\documentclass[twocolumn]{aastex631}
\usepackage{CJK}
\usepackage{amsmath}
\usepackage{bm}
\usepackage{multirow,amsmath,array,booktabs}
\usepackage{xcolor}
\usepackage{ulem}

\usepackage{multirow}
\usepackage{makecell}
\usepackage{booktabs}
\usepackage{array}
\usepackage{hyperref}
\usepackage{slashed,mathtools}

\newcommand{\bbm}{\begin{bmatrix}}
\newcommand{\ebm}{\end{bmatrix}}
\newcommand{\bBm}{\begin{Bmatrix}}
\newcommand{\eBm}{\end{Bmatrix}}
\newcommand{\bpm}{\begin{pmatrix}}
\newcommand{\epm}{\end{pmatrix}}

\newcommand{\nc}{\newcommand}       
\nc{\vc}[1] {\mbox{\boldmath $#1$}} 
\nc{\del}       {\partial}              
\nc{\bra}       {\langle}               
\nc{\ket}       {\rangle}               
\nc{\bras}[1]   {\langle #1|}           
\nc{\kets}[1]   {|#1\rangle}            
\nc{\mapleft}[1]{           
	\smash{\mathop{\,          %
			\hbox to 1.5cm{\rightarrowfill}\, }\limits_{#1}}}
\nc{\beq}     {\begin{eqnarray}} \nc{\eeq}    {\end{eqnarray}}
\nc{\nn}      {\\\nonumber} \nc{\vs}      {\vspace{-0.275cm}}
\nc{\fra}    {\frac{1}{2}}
\nc{\mb}        {\mathbf}

\graphicspath{{./}{figures/}}

\begin{document}
\begin{CJK*}{UTF8}{gbsn}

\title{Hyper-neutron stars from an {\em ab initio} calculation}

\author{Hui Tong}    
\email{htong@uni-bonn.de}
\affiliation{Helmholtz-Institut f\"{u}r Strahlen- und Kernphysik and Bethe Center for Theoretical Physics, Universit\"{a}t Bonn, D-53115 Bonn, Germany}

\author{Serdar Elhatisari}    
\email{selhatisari@gmail.com}
\affiliation{Faculty of Natural Sciences and Engineering, Gaziantep Islam Science and Technology University, Gaziantep 27010, Turkey}
\affiliation{Interdisciplinary Research Center for Industrial Nuclear Energy (IRC-INE),\\ King Fahd University of Petroleum and Minerals (KFUPM), 31261 Dhahran, Saudi Arabia}
\affiliation{Helmholtz-Institut f\"{u}r Strahlen- und Kernphysik and Bethe Center for Theoretical Physics, Universit\"{a}t Bonn, D-53115 Bonn, Germany}

\author{Ulf-G.~Mei{\ss}ner}    
\email{meissner@hiskp.uni-bonn.de}
\affiliation{Helmholtz-Institut f\"{u}r Strahlen- und Kernphysik and Bethe Center for Theoretical Physics, Universit\"{a}t Bonn, D-53115 Bonn, Germany}
\affiliation{Institute~for~Advanced~Simulation~(IAS-4),~Forschungszentrum~J\"{u}lich,~D-52425~J\"{u}lich,~Germany}
\affiliation{Center for Advanced Simulation and Analytics (CASA),~Forschungszentrum~J\"{u}lich,~D-52425~J\"{u}lich,~Germany}
\affiliation{Peng Huanwu Collaborative Center for Research and Education,  Beihang University, Beijing 100191, China}

\begin{abstract}
The equation of state (EoS) of neutron matter plays a decisive role to understand the neutron star properties and the gravitational waves from neutron star mergers. At sufficient densities, the appearance of hyperons generally softens the EoS, leading to a reduction in the maximum mass of neutron stars well below the observed values of about 2 solar masses. Even though repulsive three-body forces are known to solve this so-called ``hyperon puzzle'', so far performing \textit{ab initio} calculations with a substantial number of hyperons for neutron star properties has remained elusive. Starting from the newly developed auxiliary field quantum Monte Carlo algorithm to simulate hyper-neutron matter (HNM) without any sign oscillations, we derive three distinct EoSs by employing the state-of-the-art Nuclear Lattice Effective Field Theory. We include $N\Lambda$, $\Lambda\Lambda$ two-body forces, $NN\Lambda$, and $N\Lambda\Lambda$ three-body forces. Consequently, we determine essential astrophysical quantities such as the neutron star mass, radius, tidal deformability, and the universal $I$-Love-$Q$ relation. The maximum mass, radius and tidal deformability of a $1.4M_\odot$ neutron star are predicted to be $2.17(1)(1)~M_\odot$, $R_{1.4M\odot}=13.10(1)(7)$~km, and $\Lambda_{1.4M_\odot}=597(5)(18)$, respectively, based on our most realistic EoS. These predictions are in good agreement with the latest astrophysical constraints derived from observations of massive neutron stars, gravitational waves, and joint mass-radius measurements. Also, for the first time in \textit{ab initio} calculations, we investigate both non-rotating and rotating neutron star configurations. The results indicate that the impact of rotational dynamics on the maximum mass is small, regardless of whether hyperons are present in the EoS or not.
\end{abstract}


\section{Introduction}\label{Introduction}

Neutron stars arguably emerge as the most captivating and enigmatic astrophysical objects in the era of multi-messenger astronomy~\citep{LIGOScientific:2017vwq,LIGOScientific:2018cki,Tong2020PRC,Huth:2021bsp,MUSES:2023hyz,Tsang:2023vhh,Marino:2024gpm}.
They are composed of the densest form of baryonic matter observed in the universe, and their interiors may harbor exotic and previously unknown forms of matter~\citep{Lattimer:2004pg,Gal:2016boi,Burgio:2021vgk}.
The recent various neutron star observations, including gravitational waves, electromagnetic radiation, and X-ray bursts, have opened new frontiers for studying the neutron star properties.
These observations are expected to provide crucial insights into the mysterious dense matter at the core of neutron stars, illuminating the fundamental interactions and behavior of matter at supra-saturation nuclear densities.

The recent precise measurements of neutron star masses offer valuable constraints on the equation of state (EoS) of neutron star matter~\citep{Demorest2010Nature,Antoniadis2013,Fonseca2016APJ,NANOGrav:2017wvv,Cromartie2020NA,Fonseca2021APJ}, which is essential for refining theoretical models of their internal composition and behavior.
However, the discovery of neutron stars with masses exceeding 2 solar masses challenges many previous theoretical predictions involving exotic non-nucleonic components, such as hyperons.
This has led to the emergence of the "hyperon puzzle," a long-standing issue in the field of nuclear physics and astrophysics~\citep{Schaffner-Bielich:2008zws,Takatsuka:2008zz,Djapo:2008au,Vidana:2010ip,Schulze:2011zza,Weissenborn:2011ut,Yamamoto:2013ada,Astashenok:2014pua,Lonardoni:2014bwa,Maslov:2015msa,Chatterjee:2015pua,Masuda:2015kha,Haidenbauer:2016vfq,Logoteta:2019utx,Gerstung:2020ktv,Friedman:2022bpw}.
The appearance of hyperons at higher densities typically softens the EoS, resulting in a reduction in the maximum mass of neutron stars, which conflicts with observations of massive neutron stars.
Resolving this puzzle is crucial not only for advancing our understanding of neutron star physics but also for comprehending the complex interplay between strong nuclear forces and the behavior of dense matter under extreme conditions, including the onset of hyperons, kaons, or other exotic particles.
Furthermore, neutron stars, due to their extreme compactness, exhibit remarkably high rotational speeds compared to other astrophysical objects.
PSR J1748-2446ad, with a rotational frequency of 716 Hz, is the fastest known pulsar~\citep{Hessels2006Science}. At such extreme speeds, neutron stars experience significant centrifugal forces, leading to an oblate shape that deviates from spherical symmetry. It can be modeled as axisymmetric, rigidly rotating bodies under Einstein's general relativity.
These rapid rotations give rise to unique phenomena that are pivotal for imposing tighter constraints on the EoS~\citep{Komatsu1989MNRAS,Weber1991PLB,Stergioulas1995APJ,Glendenning1997PRL,Hessels2006Science,Li2016PRD,Qu:2024duu}.

On the theoretical side, the EoS can be derived through various nuclear many-body theories~\citep{Oertel2017,BURGIO2021PPNP,Tong:2022yml,Sedrakian2023PPNP}.
Especially, \textit{ab initio} methods using realistic nucleon-nucleon (NN) interactions stand out for their predictive power, free from the uncertainties of adjustable parameters.
Among these, the state-of-the-art Nuclear Lattice Effective Field Theory (NLEFT) combines the theoretical principles of effective field theory (EFT) with advanced numerical techniques, offering a powerful approach to solving quantum many-body systems~\citep{Lee:2008fa,Lahde:2019npb}.
This method has been used to describe the atomic nuclei~\citep{Borasoy:2005yc} and neutron matter~\citep{Lee:2004qd} in pionless EFT at leading order (LO), the Hoyle state in $^{12}$C~\citep{Epelbaum:2011md} and $\alpha$-$\alpha$ scattering~\citep{Elhatisari:2015iga} in chiral EFT at next-to-next-to-leading order (N2LO).
More recently, it was extended to the properties of atomic nuclei and the EoS of neutron and symmetric nuclear matter in chiral EFT at next-to-next-to-next-to-leading order (N3LO)~\citep{Elhatisari:2022zrb}. In addition, this method has been used to formulate an EFT with only four parameters and built on Wigner's SU(4) spin-isospin symmetry~\citep{Wigner:1936dx}.
This EFT can reproduce light- and medium-mass nuclei and neutron matter EoS with percent-level accuracy~\citep{Lu:2018bat}. Applications include nuclear thermodynamics~\citep{Lu:2019nbg}, cluster studies in hot dilute matter~\citep{Ren:2023ued}, the geometry of $^{12}$C states~\citep{Shen:2021kqr,Shen:2022bak}, and resolving the alpha-particle monopole transition form factor puzzle~\citep{Meissner:2023cvo}.
The first exploration of $\Lambda N$ scattering was performed on the lattice in Ref.~\citep{ShahinMSc}.
The $\Lambda$ particle was included into the NLEFT framework in Ref.~\citep{Frame:2020mvv} using the impurity lattice Monte Carlo (ILMC) method~\citep{Elhatisari:2014lka}.
This study focused on calculating the binding energies of light hypernuclei, specifically ${}_{\Lambda}^{3}$H, ${}_{\Lambda}^{4}$H, and ${}_{\Lambda}^{5}$He.
The ILMC method was also extended to study the systems containing two impurities~\citep{Hildenbrand:2022imw}.
Recently, a novel auxiliary field quantum Monte Carlo (AFQMC) algorithm was introduced to efficiently investigate hyper-neutron systems with an arbitrary number of hyperons~\citep{TONG2025}.
Based on the achievements of EFT within Wigner's SU(4) spin-isospin symmetry, referred to as the minimal nuclear interaction, and the newly developed AFQMC algorithm for hyper nuclear systems without any sign oscillations~\citep{TONG2025}, we employ pionless EFT at LO for nucleons~\citep{Konig:2016utl} throughout this work. This approach leverages the minimal nuclear interaction as a foundation for our hyper-neutron matter (HNM) EoS calculations.
We also utilize minimal interactions for the hyperon-nucleon and hyperon-hyperon interactions.

In this work, it is timely and interesting to study the EoS and the properties of non-rotating and rotating neutron stars with hyperons from the NLEFT.
This paper is arranged as follows.
The theoretical framework of NLEFT, the EoS, and neutron star properties are briefly introduced in Section~\ref{Theoreticalframework}.
In Section~\ref{Resultsdiscussions}, the properties of neutron stars and related discussions are presented.
The summary is given in Section~\ref{summary}.

\section{Theoretical framework}\label{Theoreticalframework}

\subsection{Nuclear Lattice Effective Field Theory}\label{sec:NLEFT}

 We map the four-dimensional space-time on a finite volume with spatial length $L$ in all three
 directions and temporal length $L_t$ in the Euclidean time direction. We further discretize
 the space and time directions in terms of a spatial and temporal lattice spacing $a$ and $a_t$, respectively.
 Our basic degrees of freedom are nucleons, so as not to resolve their inner structure, $a \gtrsim 1\,$ is required.
 The spatial coordinates on the lattice are given by a three-vector $\vec{n} =(n_x,n_y,n_z)$ with
 $n_x,n_y,n_z$ integers.
 The temporal lattice spacing is usually taken much smaller. For more details, the reader is referred to
 Ref.~\citep{Lahde:2019npb}.

The Hamiltonian for hyper-nuclear system is defined as,
\begin{align}
 H= & H_{\rm free}+\frac{c_{NN}}{2}\sum_{\vec{n}}
\,:\,\left[
\tilde{\rho}(\vec{n})
\right]^2
\,:\,
+\frac{c_{NN}^{T}}{2}\sum_{I,\vec{n}}
\,:\,
\left[
\tilde{\rho}_{I}(\vec{n})
\right]^2
\,:\,
\nonumber\\
& + c_{N\Lambda}\sum_{\vec{n}}
\,:\,
\tilde{\rho}(\vec{n})
\tilde{\xi}(\vec{n})
\,:\,
+ \frac{c_{\Lambda\Lambda}}{2}\sum_{\vec{n}}
\,:\,
\left[
\tilde{\xi}(\vec{n})
\right]^2
\,:\,
\nonumber\\
 &
+V^{\rm GIR}_{NN}
+V^{\rm GIR}_{N\Lambda}
+V^{\rm GIR}_{\Lambda\Lambda}
+V_{\rm Coulomb}
\nonumber\\
 &
+V_{NNN}
+V_{NN\Lambda}
+V_{N\Lambda\Lambda}
\,,
\label{eq:H-001}
\end{align}
where $H_{\rm free}$ is the kinetic energy term defined by using fast Fourier transforms to produce the exact dispersion relations $E_N = p^2/(2m_{N})$ and $E_\Lambda =p^2/(2m_{\Lambda})$ with nucleon mass $m_{N}=938.92$~MeV and hyperon mass $m_{\Lambda}=1115.68$~MeV, respectively, the $::$ symbol indicates normal ordering, $c_{NN}$ is the coupling constant of the SU(4) symmetric short-range two-nucleon interaction, $c_{NN}^{T}$ is the coupling constant of the isospin-dependent short-range two-nucleon interaction, that breaks SU(4) symmetry, $c_{N\Lambda}$ ($c_{\Lambda\Lambda}$) is the coupling constant of the spin-symmetric short-ranged hyperon-nucleon (hyperon-hyperon) interaction, and $\tilde{\rho}$ ($\tilde{\xi}$) is nucleon (hyperon) density operator, that is smeared both locally and non-locally~\citep{Elhatisari:2016owd},
\begin{widetext}
\begin{align}
\tilde{\rho}(\vec{n}) = \sum_{i,j=0,1}
\tilde{a}^{\dagger}_{i,j}(\vec{n}) \, \tilde{a}^{\,}_{i,j}(\vec{n})
+
s_{\rm L}
 \sum_{|\vec{n}-\vec{n}^{\prime}|^2 = 1}
 \,
 \sum_{i,j=0,1}
\tilde{a}^{\dagger}_{i,j}(\vec{n}^{\prime}) \, \tilde{a}^{\,}_{i,j}(\vec{n}^{\prime})
\,,
\end{align}
\begin{align}
\tilde{\rho}_{I}(\vec{n})= \sum_{i,j,j^{\prime}=0,1}
\tilde{a}^{\dagger}_{i,j}(\vec{n}) \,\left[\tau_{I}\right]_{j,j^{\prime}} \, \tilde{a}^{\,}_{i,j^{\prime}}(\vec{n})
+
s_{\rm L}
 \sum_{|\vec{n}-\vec{n}^{\prime}|^2 = 1}
 \,
  \sum_{i,j,j^{\prime}=0,1}
\tilde{a}^{\dagger}_{i,j}(\vec{n}^{\prime}) \,\left[\tau_{I}\right]_{j,j^{\prime}} \, \tilde{a}^{\,}_{i,j^{\prime}}(\vec{n}^{\prime})\,,
\end{align}
\begin{align}
\tilde{\xi}(\vec{n}) = \sum_{i=0,1}
\tilde{b}^{\dagger}_{i}(\vec{n}) \, \tilde{b}^{\,}_{i}(\vec{n})
+
s_{\rm L}
 \sum_{|\vec{n}-\vec{n}^{\prime}|^2 = 1}
 \,
 \sum_{i=0,1}
\tilde{b}^{\dagger}_{i}(\vec{n}^{\prime}) \, \tilde{b}^{\,}_{i}(\vec{n}^{\prime})
\,.
\end{align}
\end{widetext}
The smeared annihilation and creation operators, $\tilde{a}$ ($\tilde{b}$) and $\tilde{a}^{\dagger}$ ($\tilde{b}^{\dagger}$) for nucleons (hyperons), have with spin $i = 0, 1$ (up, down) and isospin $j = 0, 1$ (proton, neutron) indices,
\begin{align}
\tilde{a}_{i,j}(\vec{n})=a_{i,j}(\vec{n})+s_{\rm NL}\sum_{|\vec{n}^{\prime}-\vec{n}|=1}a_{i,j}(\vec{n}^{\prime})\,,
\label{eqn:rho-local-nonlocal}
\end{align}
\begin{align}
\tilde{b}_{i}(\vec{n})=b_{i}(\vec{n})+s_{\rm NL}\sum_{|\vec{n}^{\prime}-\vec{n}|=1}b_{i}(\vec{n}^{\prime}).
\label{eqn:xi-local-nonlocal}
\end{align}
In Eq.~(\ref{eq:H-001}), $V_{\text{Coulomb}}$ represents the Coulomb interaction~\citep{Li:2018ymw}. $V^{\text{GIR}}_{NN}$, $V^{\text{GIR}}_{N\Lambda}$, and $V^{\text{GIR}}_{\Lambda\Lambda}$, denote the Galilean invariance restoration (GIR) interactions for the nucleon-nucleon, nucleon-hyperon, and hyperon-hyperon interactions, respectively~\citep{Li:2019ldq}.
$V_{NNN}$, $V_{NN\Lambda}$, and $V_{N\Lambda\Lambda}$ are the three-baryon interactions.
The three-baryon interactions are defined with two different choices of local smearing,
\begin{align}\label{eq:VNNN}
V_{NNN}
=
\sum_{i = 1,2}
\frac{c_{NNN}^{(d_i)}}{6}
\,
\sum_{\vec{n}}
\,:\,
\left[
\rho^{(d_i)}(\vec{n})
\right]^3
\,:\,,
\end{align}
where the parameter $d_i$ denotes the range of local smearing with $0 \leq d_1 < d_2 \leq 3$ (in lattice units). Similarly, the three-baryon interactions consisting of two nucleons and one hyperon are defined with two different choices of local smearing,
\begin{align}
V_{NN\Lambda}
=
\sum_{i = 1,2}
\frac{c_{NN\Lambda}^{(d_i)}}{2}
\,
\sum_{\vec{n}}
\,:\,
\left[
\rho^{(d_i)}(\vec{n})
\right]^2 \xi^{(d_i)}(\vec{n})
\,:\,,
\label{eqn:V-NNY}
\end{align}
and the interactions involving one nucleon and two hyperons are expressed by also two different choices of local smearing,
\begin{align}
V_{N\Lambda\Lambda}
=
\sum_{i = 1,2}
\frac{c_{N\Lambda\Lambda}^{(d_i)}}{2}
\,
\sum_{\vec{n}} \,
 \,:\,
 \rho^{(d_i)}(\vec{n})  \,
\left[
\xi^{(d_i)}(\vec{n})
\right]^2
\,:\,,
\label{eqn:V-NYY}
\end{align}
where ${\rho}$ (${\xi}$) is then purely locally smeared nucleon (hyperon) density operator with annihilation and creation operators, ${a}$ (${b}$) and ${a}^{\dagger}$ (${b}^{\dagger}$) for nucleons (hyperons),
\begin{align}
  \begin{split}
    {\rho}^{(d)}(\vec{n}) &= \sum_{i,j=0,1}
    {a}^{\dagger}_{i,j}(\vec{n}) \, {a}^{\,}_{i,j}(\vec{n})\\
    &+s^{\rm 3B}_{\rm L}\sum_{|\vec{n}-\vec{n}^{\prime}|^2 = 1}^{d}
     \,
    \sum_{i,j=0,1}
    {a}^{\dagger}_{i,j}(\vec{n}^{\prime}) \, {a}^{\,}_{i,j}(\vec{n}^{\prime})
     \,,
    \label{eqn:rho-local}
  \end{split}
\end{align}
\begin{align}
  \begin{split}
    {\xi}^{(d)}(\vec{n}) &= \sum_{i=0,1}
    {b}^{\dagger}_{i}(\vec{n}) \, {b}^{\,}_{i}(\vec{n})\\
    &+
    s^{\rm 3B}_{\rm L}
    \sum_{|\vec{n}-\vec{n}^{\prime}|^2 = 1}^{d}
    \,
    \sum_{i=0,1}
    {b}^{\dagger}_{i}(\vec{n}^{\prime}) \, {b}^{\,}_{i}(\vec{n}^{\prime})
    \,.
  \label{eqn:xi-local}
  \end{split}
\end{align}
Here, the parameter $d$ gives the range of local smearing, and $s^{\rm 3B}_{\rm L}$ defines the strength of the local smearing.

In our lattice simulations, we use the AFQMC method, which effectively suppresses sign oscillations.
The following discussion begins with a discrete auxiliary field formulation for the SU(4) symmetric short-ranged two-nucleon interaction given in Eq.~(\ref{eq:H-001}),
\begin{align}
   : \exp \left( -\frac{a_{t} \, c_{NN}}{2} \, \tilde{\rho}^2
          \right) :
=
\sum_{k = 1}^{3} \, w_{k} \,
: \exp \left( \sqrt{-a_{t} \, c_{NN}}  \, s_{k} \, \tilde{\rho} \right) \, :
\label{eqn:AFQMC-NN}
\end{align}
where $a_{t}$ is the temporal lattice spacing. From a Taylor expansion of Eq.~(\ref{eqn:AFQMC-NN}), we determine the constants $s_{k}$ and $w_k$ as $s_{1} = -s_{3}=\sqrt{3}$, $s_{2} = 0$, $w_{1} = w_3 = 1/6$ and $w_2 = 2/3$.
Since we use minimal interactions for the hyperon-nucleon and hyperon-hyperon interactions, the spin and isospin independent two-baryon interaction in Eq.~(\ref{eq:H-001}) is expressed as,
\begin{align}
\begin{split}
V_{\rm 2B} &= \frac{c_{NN}}{2}\sum_{\vec{n}}
\,:\,
\left[
\tilde{\rho}(\vec{n})
\right]^2
\,:\,
+ c_{N\Lambda}\sum_{\vec{n}}
\,:\,
\tilde{\rho}(\vec{n})
\tilde{\xi}(\vec{n})
\,:\,\\
&+ \frac{c_{\Lambda\Lambda}}{2}\sum_{\vec{n}}
\,:\,
\left[
\tilde{\xi}(\vec{n})
\right]^2
\,:\,
\,,
\end{split}
\label{eqn:NY-Potential-000}
\end{align}
and it can be rewritten in the following form,
\begin{align}
\begin{split}
V_{\rm 2B} &= \frac{c_{NN}}{2}\sum_{\vec{n}}
\,:\,
\left[
\tilde{\slashed{\rho}}(\vec{n})
\right]^2
\,:\,\\
&+
\frac{1}{2}
\left(
c_{\Lambda\Lambda}
-\frac{c_{N\Lambda}^2}{c_{NN}}
\right)
\sum_{\vec{n}}
\,:\,
\left[
\tilde{\xi}(\vec{n})
\right]^2
\,:\,
\,,
\label{eqn:NY-Potential-010}
\end{split}
\end{align}
where $\tilde{\slashed{\rho}}$ is defined as,
\begin{align}
\tilde{\slashed{\rho}} =
\tilde{\rho} +
\frac{c_{ N\Lambda}}{c_{NN}} \,  \tilde{\xi}
\,.
\label{eqn:Density-rho-bar}
\end{align}
In hypernuclear systems, the nucleon-nucleon interaction strength $c_{NN}$ is significantly stronger than the nucleon-hyperon interaction strength $c_{N\Lambda}$. As a result, the overall interaction strength of the second term in Eq.(\ref{eqn:NY-Potential-010}) is naturally weak. Consequently, we treat the first term in Eq.(\ref{eqn:NY-Potential-010}) non-perturbatively, while the second term is computed using first-order perturbation theory.

Employing a Hubbard-Stratonovich transformation for the first term in Eq.~(\ref{eqn:NY-Potential-010}) enables the simulations of systems consisting of both arbitrary number of nucleons and  arbitrary number of $\Lambda$ hyperons with a single auxiliary field,
\begin{align}
   : \exp \left( -\frac{a_{t} \, c_{NN}}{2} \, \tilde{\slashed{\rho}}^2\right) :
=
\sum_{k = 1}^{3} \, w_{k} \,
: \exp \left( \sqrt{-a_{t} \, c_{NN}}  \, s_{k} \,  \tilde{\slashed{\rho}} \right) \,. :
\label{eqn:AFQMC-NY}
\end{align}
It is evident that the solution for the auxiliary field variables $s_{k}$ and weights $w_{k}$ is consistent with systems containing only nucleons.
For the two-nucleon interaction $\sim c_{NN}^T$,
known to break SU(4) symmetry and to induce significant sign oscillations, which was previously disregarded in minimal nuclear interaction studies~\citep{Lu:2018bat,Lu:2019nbg,Ren:2023ued,Shen:2021kqr,Shen:2022bak,Meissner:2023cvo}, we also employ a Hubbard-Stratonovich transformation and introduce a discrete auxiliary field. Note that this term is required to obtain
a good description of nuclear matter as discussed below.

\subsection{Neutron star EoS and Neutron star properties}

Hyper-neutron matter consists of neutrons and a fraction of $\Lambda$ hyperons defined as $x_\Lambda=\rho_\Lambda/\rho$, where $\rho=\rho_N+\rho_\Lambda$ represents the total baryon density of the system. Therefore, the neutron and hyperon densities are written as $\rho_N=(1-x_\Lambda)\rho$ and $\rho_\Lambda=\rho x_\Lambda$, respectively. The HNM energy per particle can be expressed as
\begin{equation}
e_{\rm HNM}(\rho,x_\Lambda) = \frac{E_{\rm HNM}(\rho,x_\Lambda)}{N_{\rm tot}} + m_N(1-x_\Lambda) + m_\Lambda x_\Lambda,
\end{equation}
where $E_{\rm HNM}(\rho,x_\Lambda)$ and $N_{\rm tot}=N_N+N_\Lambda$ denotes the total energy of HNM and the total number of baryons.
Now, our objective is to compute $e_{\scriptscriptstyle \rm HNM}(\rho,x_\Lambda)$, and subsequently,
calculate the energy density $\varepsilon_{\scriptscriptstyle \rm HNM}$,
defined as $\varepsilon_{\scriptscriptstyle \rm HNM} = \rho e_{\scriptscriptstyle \rm HNM}$. The chemical potentials for neutrons and hyperons, denoted by $\mu_N(\rho,x_\Lambda)$ and
$\mu_\Lambda(\rho,x_\Lambda)$ respectively, are then evaluated using the expressions,
\begin{equation}
    \mu_N(\rho,x_\Lambda)=\frac{\partial \varepsilon_{\scriptscriptstyle \rm HNM}}{\partial \rho_N},~~
    \mu_\Lambda(\rho,x_\Lambda)=\frac{\partial \varepsilon_{\scriptscriptstyle \rm HNM}}{\partial \rho_\Lambda}\,.
\end{equation}
The hyperon fraction as a function of the baryon density, $x_\Lambda(\rho)$, is determined by
imposing the condition $\mu_\Lambda=\mu_N$, which yields the threshold density $\rho_\Lambda^{\rm th}$ which is
marking the point at which $x_\Lambda(\rho)$ first deviates from zero. Finally, the pressure $P(\rho)$ of HNM is obtained from the energy density,
\begin{equation}\label{ns-eos}
    P(\rho)=\rho^2\frac{d}{d\rho}\frac{\varepsilon_{\scriptscriptstyle \rm HNM}}{\rho}=\sum_{i=N,\Lambda}\rho_i\mu_i-\varepsilon_{\scriptscriptstyle \rm HNM}.
\end{equation}

Once the EoS of pure neutron matter (PNM) and HNM in the form $P(\varepsilon)$ is obtained in Eq.~\eqref{ns-eos}, the mass and radius of a neutron star can be described by the Tolman-Oppenheimer-Volkoff (TOV) equations~\citep{Tolman:1939jz,Oppenheimer:1939ne}
\begin{subequations}\label{ns-toveq}
  \begin{align}
    \frac{dP(r)}{dr}=&\ -\frac{[P(r)+\varepsilon(r)][M(r)+4\pi r^3P(r)]}{r[r-2M(r)]}, \\
	\frac{dM(r)}{dr} =&\ 4\pi r^2\varepsilon(r),
  \end{align}
\end{subequations}
where $P(r)$ is the pressure at radius $r$ and $M(r)$ is the total mass inside a sphere of radius $r$.
Furthermore, to solve the TOV equations, the EoS must cover the entire structure of the neutron star, from the crust to the core. In this work, we mainly focus on discussing the core region within the NLEFT. For the crust, we adopt the well-established EoSs formulated by Baym, Pethick, Sutherland (BPS)~\citep{Baym1971-BPS} and by Baym, Bethe, and Pethick (BBP)~\citep{Baym1971-BBP}.

Besides the masses and radii, another important property of neutron star, the tidal deformability $\Lambda$, is defined as
\begin{equation}
  \Lambda = \frac{2}{3}k_2 C^{-5},
\end{equation}
which represents the mass quadrupole moment response of a neutron star to the strong gravitational field induced by its companion. Further, $C=M/R$ is the compactness parameter, $M$ and $R$ are the neutron star mass and radius, and $k_2$~is the second love number
\begin{equation}\label{ns-luv}
  \begin{aligned}
    k_2=&\frac{8C^5}{5}(1-2C)^2[2-y_R+2C(y_R-1)]\\
    &\times\{6C[2-y_R+C(5y_R-8)] \\
    &+4C^3[13-11y_R+C(3y_R-2) +2C^2(1+y_R)] \\
    &+3(1-2C)^2[2-y_R+2C(y_R-1)]\ln(1-2C)\}^{-1},
  \end{aligned}
\end{equation}
where $y_R=y(R)$ can be calculated by solving the following differential equation:
\begin{equation}\label{yRequ}
  r\frac{d y(r)}{dr} + y^2(r)+y(r)F(r) + r^2Q(r)=0,
\end{equation}
with
\begin{subequations}
\begin{align}
  &F(r) = \left[1-\frac{2M(r)}{r}\right]^{-1}\times\left\{1-4\pi r^2[\varepsilon(r)-P(r)]\right\} ,\\
  \nonumber
  &Q(r) =  \left\{4\pi  \left[5\varepsilon(r)+9P(r)+\frac{\varepsilon(r)+P(r)}{\frac{\partial P}{\partial\varepsilon}(r)}\right]-\frac{6}{r^2}\right\}\\
  \nonumber
  &\times \left[1-\frac{2M(r)}{r}\right]^{-1}-\left[\frac{2M(r)}{r^2} +2\times4\pi r P(r) \right]^2 \\
  &\times \left[1-\frac{2M(r)}{r}\right]^{-2} .
\end{align}
\end{subequations}
The differential equation \eqref{yRequ} can be integrated together with the TOV equations with the boundary condition $y(0)=2$.

The moment of inertia is calculated under the slow-rotation approximation pioneered by Hartle and Thorne~\citep{Hartle:1967he,Hartle:1968si}, where the frequency $\Omega$ of a uniformly rotating neutron star is significantly lower than the Kepler frequency at the equator, $\Omega \ll \Omega_{\text{max}} \simeq \sqrt{M/R^3}$. In the slow-rotation approximation, the moment of inertia of a uniformly rotating, axially symmetric neutron star is given by the following expression~\citep{Fattoyev:2010tb}
\begin{equation}
	I = \frac{8\pi}{3} \int^R_0 r^4 e^{-\nu(r)} \frac{\bar{\omega}(r)}{\Omega} \frac{\epsilon(r)+P(r)}{\sqrt{1-2M(r)/r}} dr.
\end{equation}
The quantity $\nu(r)$ is a radially-dependent metric function and defined as
\begin{equation}
	\nu(r) = \frac{1}{2}\ln \left( 1- \frac{2M}{R}\right) - \int^R_r \frac{M(x)+4\pi x^3 P(x)}{x^2[1-2M(x)/x]}dx.
\end{equation}
The frame-dragging angular velocity $\bar{\omega}$ is usually obtained by the dimensionless relative frequency $\tilde{\omega}\equiv \bar{\omega}/\Omega$, which satisfies the following second-order differential equation:
\begin{equation}
  \frac{d}{dr}\left[ r^4 j(r)\frac{d\tilde{\omega}(r)}{dr}\right]  + 4r^3 \frac{dj(r)}{dr} \tilde{\omega}(r) = 0,
\end{equation}
where $j(r) = e^{-\nu(r)} \sqrt{1-2M(r)/r}$ for $r\leq R$.
The relative frequency $\tilde{\omega}(r)$ is subject to the following two boundary conditions
\begin{subequations}
  \begin{align}
    \tilde{\omega}'(0) =&\ 0,\\
    \tilde{\omega}(R) + \frac{R}{3}\tilde{\omega}'(R) =&\ 1.
  \end{align}
\end{subequations}
It should be noted that under the slow-rotation approximation, the moment of inertia is independent of the stellar frequency $\Omega$.

The quadrupole moment describes how much a neutron star is deformed away from sphericity due to rotation. It can be computed by numerically solving for the interior and exterior gravitational field of a neutron star in a slow-rotation~\citep{Hartle:1967he,Hartle:1968si} and a small-tidal-deformation approximation~\citep{Hinderer:2007mb,Hinderer:2009ca}. The quadrupole moment in this work is calculated by following the detailed instructions described in Ref.~\citep{Yagi:2013awa}.
To explore the universal $I$-Love-$Q$ relations, the following dimensionless quantities are introduced
\begin{equation}
	\bar{I} \equiv \frac{I}{M^3},\qquad \bar{Q} \equiv - \frac{QM}{(I\Omega)^2}.
\end{equation}

In addition, to describe the rapidly rotating and axisymmetric neutron star configurations in general relativity, we treat the stellar matter as a perfect fluid, characterized by the energy-momentum tensor:
\begin{equation}
  T^{\mu\nu}=(\varepsilon + P)u^{\mu}u^{\nu}-g^{\mu\nu}P,
\end{equation}
where $\varepsilon$, $P$, and $u^{\mu}$ are the energy density, pressure, and fluid's four-velocity, respectively.
We solve the Einstein field equations for an axisymmetric and stationary space-time with the metric
\begin{equation}
  \begin{split}
  	ds^2=&\ -e^{\gamma+\rho}dt^2+e^{2\alpha}(dr^2+r^2d\theta^2)\\
  		&\ +e^{\gamma-\rho}r^2\sin^2\theta(d\phi-\omega dt)^2,
  \end{split}
\end{equation}
where the metric potentials $\gamma, \rho, \alpha$, and $\omega$ are functions of the radial coordinates $r$ and the polar angle $\theta$. To numerically integrate the equilibrium equations, we employ the RNS code~\citep{Stergioulas1995APJ, Paschalidis2017LR} to calculate the equilibrium configurations of rotating neutron stars, determining their masses and radii for a given central energy density.

\section{NUMERICAL RESULTS AND DISCUSSION}\label{Resultsdiscussions}

The coupling constants for the NN interaction are determined by fitting to the two $S$-wave phase shifts of NN scattering. The results are $c_{{}^1{\rm S}_0}=-1.21\times10^{-7}$~MeV$^{-2}$ and $c_{{}^3{\rm S}_1}=-1.92\times10^{-7}$~MeV$^{-2}$ corresponding with the spin-singlet isospin-triplet and the spin-triplet isospin-singlet channel, which are related to the LECs given in Eq.~\eqref{eq:H-001} via $c_{NN}^{} = (3\,c_{{}^1{\rm S}_0} + c_{{}^3{\rm S}_1})/4, ~~
c_{NN}^{T} = (c_{{}^1{\rm S}_0} - c_{{}^3{\rm S}_1})/4$.
The two LECs of the three-nucleon forces given in Eq.~\eqref{eq:VNNN} are determined by fitting to the saturation properties of symmetric nuclear matter. This procedure involves considering all possible combinations three-nucleon forces with $d_1$ and $d_2$ such that $0 \leq d_1 < d_2 \leq 3$. These combinations and the corresponding energies per nucleon at the saturation density of symmetric nuclear matter are presented in Tab.~\ref{tab:VNNN}. The energy per nucleon at the saturation point is -16.90(0.02)(0.25)~MeV, where the first parentheses represent the statistical error and the second denote the theoretical uncertainty arising from different three-nucleon force combinations. This result is in good agreement with the empirical value. 
The parameters of the $N\Lambda$ and $\Lambda\Lambda$ interactions are determined by fitting them to experimental data~\citep{Sechi-Zorn:1968mao,Alexander:1968acu,Kadyk:1971tc,Hauptman:1977hr} and the $\Lambda\Lambda$ ${}^1S_0$ scattering phase shift from chiral EFT~\citep{Haidenbauer:2015zqb}.
The $NN\Lambda$ and $N \Lambda \Lambda$ forces are further constrained by the separation energies of single- and double-$\Lambda$ hypernuclei spanning systems from ${}_\Lambda^5$He to ${}_{\Lambda\Lambda}^{~~6}$Be, and the $\Lambda$ threshold densities $\rho_\Lambda^{\rm th}$ around $(2-3)\rho_0$ simultaneously, denoted as HNM(I), HNM(II) and HNM(III), respectively. In our calculations  we use a spatial lattice spacing of $a = 1.1$~fm and a temporal lattice spacing of $a_t = 0.2$~fm.
The local smearing parameter $s_{\rm L}=0.06$ and nonlocal smearing parameter $s_{\rm NL} = 0.6$.
For the three-baryon interaction, the local smearing parameter is $s^{\rm 3B}_{\rm L} = 0.06$.
We perform our calculations at different finite Euclidean time steps and extrapolate to the infinite Euclidean time limit using a single and double exponential ansatz~\citep{Lahde:2019npb}. Furthermore, for the computation of PNM and HNM energies we use lattices with a length of $6.6$~fm and impose the average twisted boundary conditions (ATBC) to efficiently eliminate finite volume effects.

\begin{table*}[htbp]
    \centering
    \caption{The coupling constants of the $NNN$ interaction for all possible combinations of $d_1$ and $d_2$ with $0 \leq d_1 < d_2 \leq 3$ in Eq.~\eqref{eq:VNNN}, along with the corresponding binding energy per nucleon at the saturation density of symmetric nuclear matter.}\label{tab:VNNN}
    \begin{tabular}{|C|C|}
    \hline
      {\rm Coupling~constants} &  {\rm Binding~energy}\\
      $c_{NNN}^{(d_i)}$~\rm (MeV$^{-5}$) & $E/A$~(\rm MeV)\\
    \hline
     c_{NNN}^{(d_1=0)}=-3.78\times 10^{-11},~c_{NNN}^{(d_2=1)}=~~$3.12\times 10^{-11}$ & $-17.42(2)$ \\
     c_{NNN}^{(d_1=0)}=~~2.11\times 10^{-12},~c_{NNN}^{(d_2=2)}=~~$4.42\times 10^{-12}$& $-16.90(2)$ \\
     c_{NNN}^{(d_1=0)}=~~7.27\times 10^{-12},~c_{NNN}^{(d_2=3)}=~~$2.09\times 10^{-12}$& $-16.63(2)$ \\
     c_{NNN}^{(d_1=1)}=~~1.55\times 10^{-12},~c_{NNN}^{(d_2=2)}=~~$4.23\times 10^{-12}$& $-16.89(2)$ \\
     c_{NNN}^{(d_1=1)}=~~4.96\times 10^{-12},~c_{NNN}^{(d_2=3)}=~~$1.78\times 10^{-12}$& $-16.71(2)$ \\
     c_{NNN}^{(d_1=2)}=~~6.01\times 10^{-12},~c_{NNN}^{(d_2=3)}=$-7.08\times 10^{-13}$& $-16.89(2)$ \\
    \hline
    \end{tabular}
\end{table*}

First, we compare our calculations with a few others that we consider as benchmarks, focusing exclusively on purely nucleonic scenarios. We predict the ground state energies of several light nuclei with $A=3 - 16$ based on our interaction, and the results are summarized in Tab.~\ref{tab:nuclei}.
These results are consistent with those reported in Ref.~\citep{Lu:2018bat}, except for $^3$H and $^4$He, which were used to constrain the $3N$ force therein.

\begin{table}[htbp]
    \caption{Calculated ground state energies of some light nuclei with $A=3 - 16$ compared to the empirical values (in MeV). The first (second) parentheses denote the statistical (systematic) error.}\label{tab:nuclei}
    \begin{tabular}{|C||C|C|}
    \hline
    \rm Nucleus   &  \rm NLEFT  &  \rm Exp.  \\
    \hline \hline
    \rm $^3$H    & $-9.21(4)(1)$  & $-8.48$  \\
    \rm $^4$He   & $-29.38(1)(4)$  &   $-28.3$ \\
    \rm $^8$Be   & $-58.38(3)(7)$  &  $-56.5$ \\
    \rm $^{12}$C & $-87.08(12)(11)$ &   $-92.2$ \\
    \rm $^{16}$O & $-121.84(28)(52)$ &  $-127.6$ \\
    \hline
    \end{tabular}
\end{table}

\begin{figure*}[htbp]
  \centering
  \includegraphics[width=0.45\textwidth]{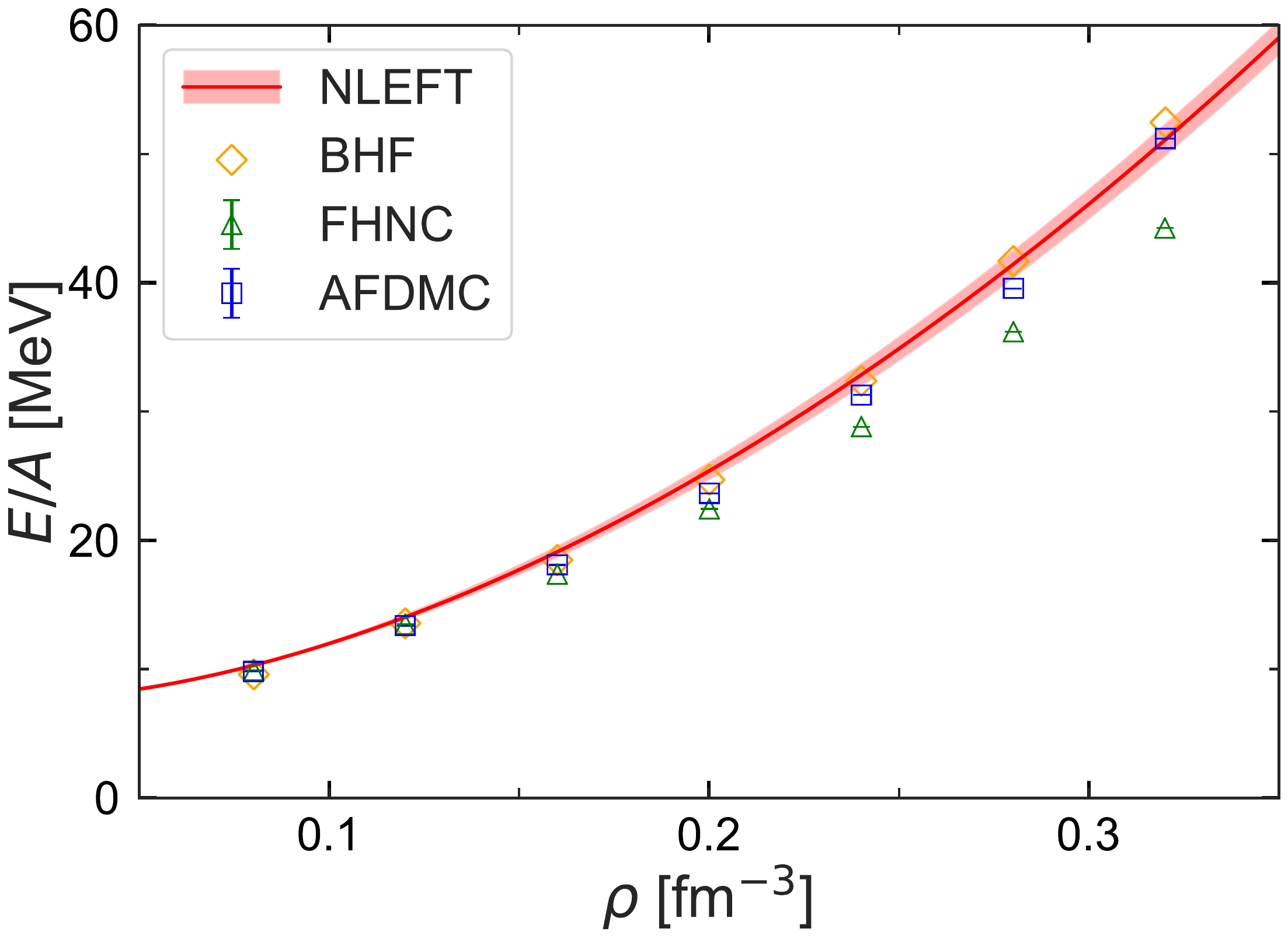}
  \hspace{0.5cm}
  \includegraphics[width=0.45\textwidth]{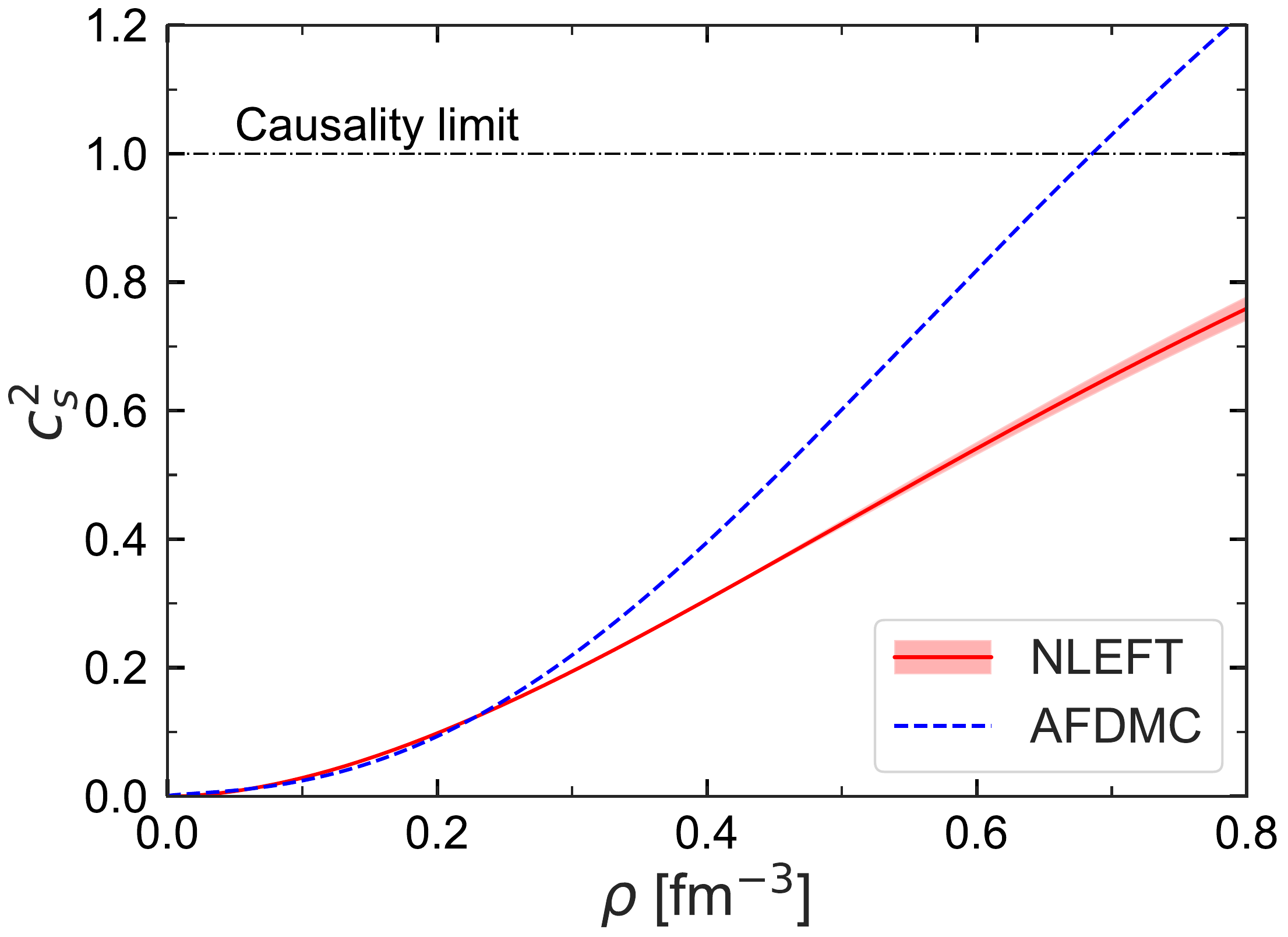}
  \caption{Left Panel: Neutron matter EoS as obtained from the NLEFT. The red shaded band represents our results with the uncertainty from three-nucleon forces and Monte Carlo errors, while the red solid curve denotes the mean value. The results obtained with other many-body methods~\citep{Lovato:2022apd} are also shown, including the Brueckner-Hartree-Fock (BHF) theory, Fermi hypernetted chain (FHNC), and Auxiliary Field Diffusion Monte Carlo (AFDMC) using the AV18 two-nucleon and Urbana~IX (UIX) three-nucleon forces. Right Panel: Speed of sound as a function of density for the pure neutron matter (PNM). The blue dashed curve is calculated with the Argonne V8' (AV8') and the UIX forces from the AFDMC~\citep{Lonardoni:2014bwa}. The dot-dashed line represents the causality limit $c_s^2=1$.
  }
  \label{fig1}
\end{figure*}

Then, we compare our results for PNM with that of Lovato et al.~\citep{Lovato:2022apd} in the left panel of Fig.~\ref{fig1}.
They presented the neutron matter EoS as derived from three independent many-body methods: Brueckner-Hartree-Fock (BHF), Fermi hypernetted chain (FHNC), and Auxiliary Field Diffusion Monte Carlo (AFDMC).
Our results are consistent with theirs where the AV18 two-nucleon and UIX three-nucleon forces were employed.
Only at densities higher than approximately 0.24~fm$^{-3}$, the energies from the FHNC method are lower than our results.
It should be emphasized that the uncertainty in our calculation for PNM is quite small.
In the right panel of Fig.~\ref{fig1}, we compare our work to the pioneering calculations of Lonardoni et al.~\citep{Lonardoni:2014bwa}.
They perform AFDMC simulations with $N_n = 38, 54, 66$ neutrons.
For the nucleonic sector, they use the phenomenological well-motivated AV8' and UIX two- and three-body forces.
Notably, their PNM EoS is stiffer compared to our results and exceeds the causality limit for the speed of sound at densities above $\rho \simeq 0.68$~fm$^{-3}$.
Although PNM is an idealized system that does not directly exist in neutron stars, it serves as a theoretical benchmark for modeling dense matter~\citep{Lovato:2022apd}. Since the neutron star EoS can exhibit stiffness trends similar to or even stiffer than those of PNM~\citep{Krastev:2006ii}, maintaining causality in the PNM EoS ensures its physical reliability and strengthens its applicability to astrophysical scenarios. A violation of causality in PNM may indicate potential issues with the underlying nuclear interactions or many-body methods, making the model unreliable at high densities.

\begin{figure}[htbp]
  \centering
  \includegraphics[width=0.47\textwidth]{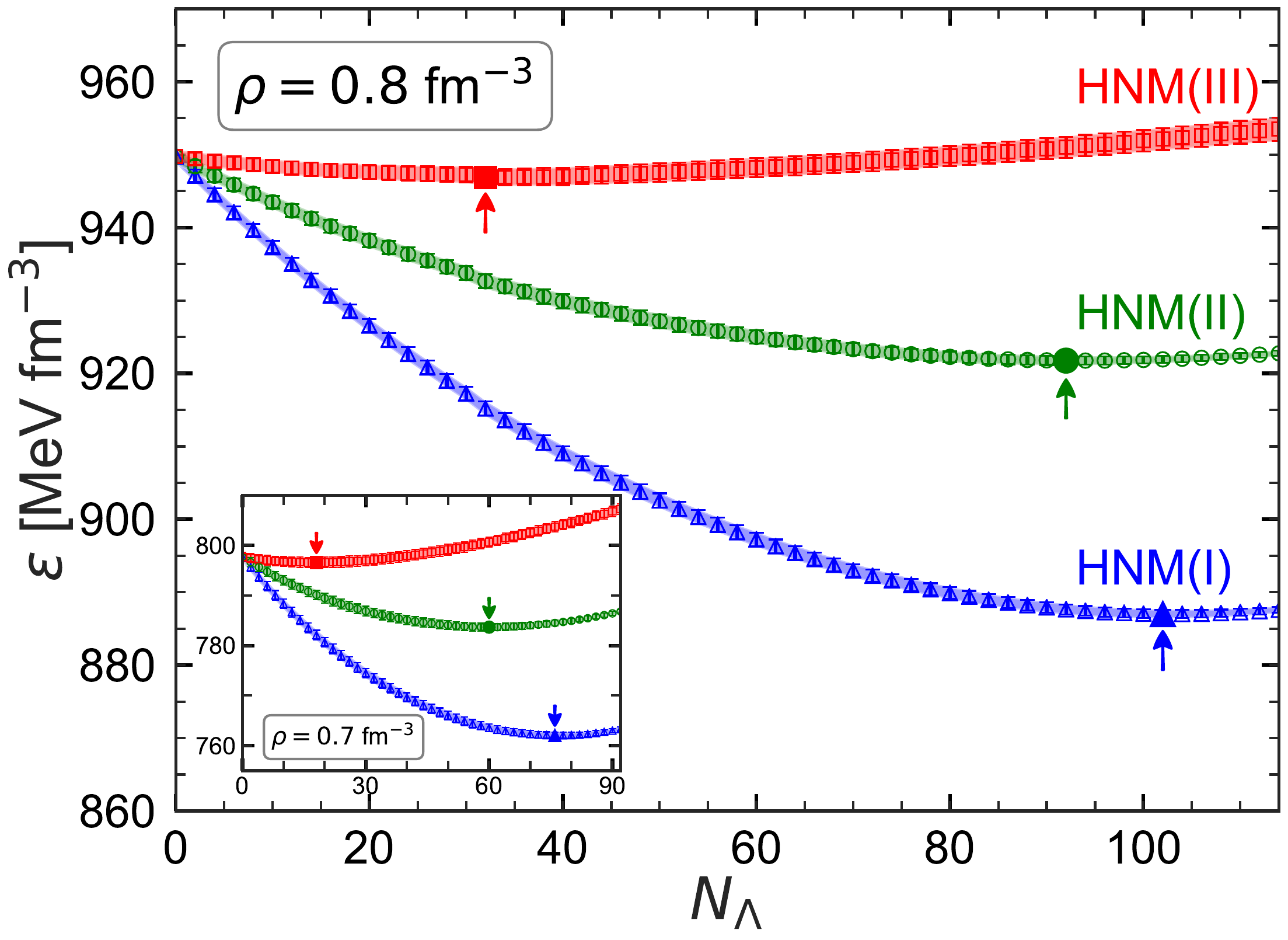}
  \caption{Energy density for hyper-neutron matter (HNM). The energy density $\varepsilon$ as a function of different numbers for $\Lambda$ hyperons is shown for densities $\rho=0.8$~fm$^{-3}$ and $0.7$~fm$^{-3}$ (inset). The blue triangles, green circles, and red squares represent the energy density $\varepsilon$ of HNM with hyperons interacting via the two-body interactions and the three-body interactions. The differences between HNM(I), HNM(II) and HNM(III) are the three-body $NN\Lambda $ and $N\Lambda\Lambda$ interactions. The shaded regions represent the uncertainty from the three-baryon forces and Monte Carlo errors. The arrows and the solid triangle, circle, and square denote the lowest energy density.}
  \label{fig2}
\end{figure}

The energy density $\varepsilon$ by using the two-body interactions ($NN,N\Lambda,\Lambda\Lambda$) and the three-body interactions ($NNN,NN\Lambda,N\Lambda\Lambda$) are shown in Fig.~\ref{fig2} for different numbers of $\Lambda$ hyperons. The differences between HNM(I), HNM(II), and HNM(III) are the three-body $NN\Lambda$ and $N\Lambda\Lambda$ interactions. The shaded regions represent the uncertainty from the three-baryon forces and Monte Carlo errors. The given density of $\rho=0.8$~fm$^{-3}$, which is about five times the empirical nuclear matter saturation density, $\rho_0$, can be encountered in the core of a neutron star. It should be noted that the quantity of $\Lambda$ hyperons corresponding to the lowest energy density is intricately linked to accurately determining the chemical equilibrium conditions. In contrast to the groundbreaking study~\citep{Lonardoni:2014bwa} where the number of $\Lambda$ hyperons was varied from 1 to 14, the present study indicates that the number of required $\Lambda$ hyperons is comparable to the number of neutrons, especially at higher densities. For instance, as depicted in Fig.~\ref{fig2}, to fulfill the equilibrium condition $\mu_N=\mu_\Lambda$ at $\rho=0.8$~fm$^{-3}$, 102, 92, and 32 $\Lambda$ hyperons are required to obtain the lowest energy density for HNM(I), HNM(II), and HNM(III), respectively. Similarly, at $\rho = 0.7$~fm$^{-3}$, 76, 60, and 18 hyperons are needed for the same purpose in HNM(I), HNM(II), and HNM(III), in order.

\begin{figure}[htbp]
  \centering
  \includegraphics[width=0.47\textwidth]{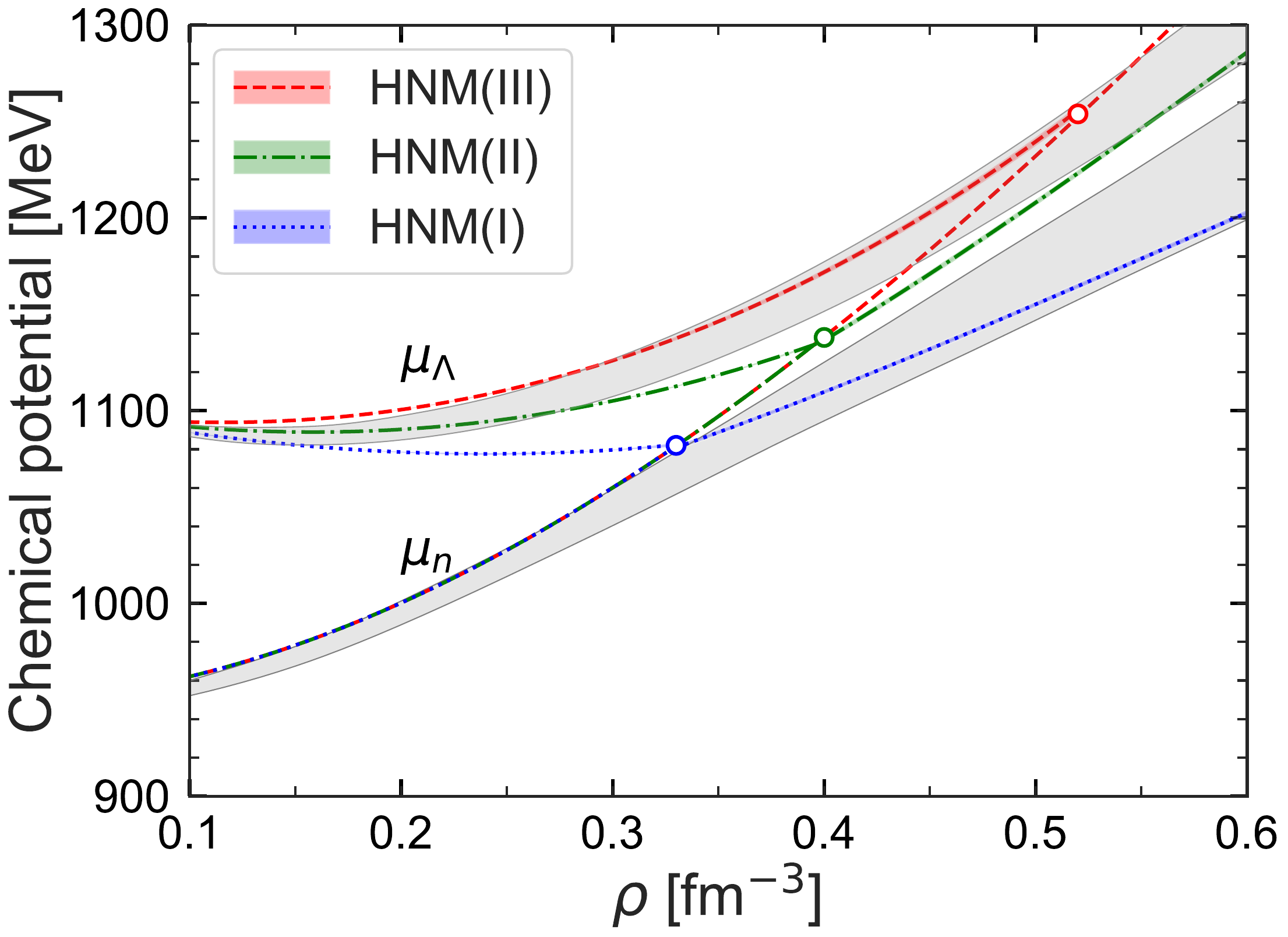}\\
  \caption{The chemical potentials for neutrons and $\Lambda$ hyperons. The $\Lambda$ threshold densities $\rho_\Lambda^{\rm th}$ are marked by open circles. The chemical equilibrium conditions, $\mu_\Lambda = \mu_n$, are fulfilled above $\rho_\Lambda^{\rm th}$. The gray shaded area indicates the values by using the chiral SU(3) interactions NLO19 with two and three-body forces $(N\Lambda + NN\Lambda)$~\citep{Gerstung:2020ktv}.}
  \label{fig3}
\end{figure}

The chemical potentials for neutrons and $\Lambda$ hyperons are shown in Fig.~\ref{fig3}.
When the density is below the threshold value, the chemical potential of the $\Lambda$ hyperon exceeds that of the neutron.
Once the density reaches and is above the threshold value, the chemical potentials of both particles become equal.
Moreover, HNM~(III) gives the largest chemical potential among our three HNM systems.
We also compare our work with that of Gerstung et al.~\citep{Gerstung:2020ktv}.
For the $\Lambda N$ interaction, they consider two next-to-leading order chiral
EFT representations, called NLO13~\citep{Haidenbauer:2013oca} and NLO19~\citep{Haidenbauer:2019boi}.
For the three-body forces, they use the leading $\Lambda NN$ representation
based on chiral EFT (contact terms, one-pion and two-pion exchanges) with the inclusion of the $\Lambda NN \leftrightarrow \Sigma NN$ transition~\citep{Petschauer:2015elq} in an effective density-dependent two-body approximation~\citep{Petschauer:2016pbn}.
The pertinent LECs are given in terms of decuplet resonance saturation and leave one
with  two $B^*BBB$ couplings, where $B$ denotes the baryon octet and $B^*$ the decuplet.
If one only considers the $\Lambda NN$ force, these two LECs appear in the
combination $H'=H_1+H_2$. No $\Lambda\Lambda N$ force was considered in~\citep{Gerstung:2020ktv}. The LECs $H_1$ and $H_2$ were constrained in Ref.~\citep{Gerstung:2020ktv} so that
the $\Lambda$ single-particle potential in infinite matter is $U_\Lambda (\rho \simeq \rho_0) =-30$~MeV
\citep{Gal:2016boi}. Due to numerical instabilities in calculation of the Brueckner $G$-matrix, the computation  can only be done up to
densities $\rho \simeq 3.5\rho_0$. The authors of Ref.~\citep{Gerstung:2020ktv} then use a
quadratic polynomial to extrapolate to higher densities. They calculate the chemical potential for the neutrons and $\Lambda$s from the Gibbs-Duhem relation using a microscopic EoS computed from a chiral nucleon-meson field theory in combination with functional renormalization group methods. The parameter combinations $(H_1,~H_2)$ were chosen so that the $\Lambda$ single-particle potential becomes maximally repulsive at higher densities. The resulting chemical potentials are displayed in Fig.~\ref{fig3} for the NLO19 $\Lambda N$ forces. These agree well with the HNM(III) chemical potentials up to $\rho \simeq 2.5\rho_0$ but show, different to what we find, no crossing. Note that the forces discussed in Ref.~\citep{Gerstung:2020ktv} have
not been applied to finite nuclei.

\begin{figure}[htbp]
  \centering
  \includegraphics[width=0.47\textwidth]{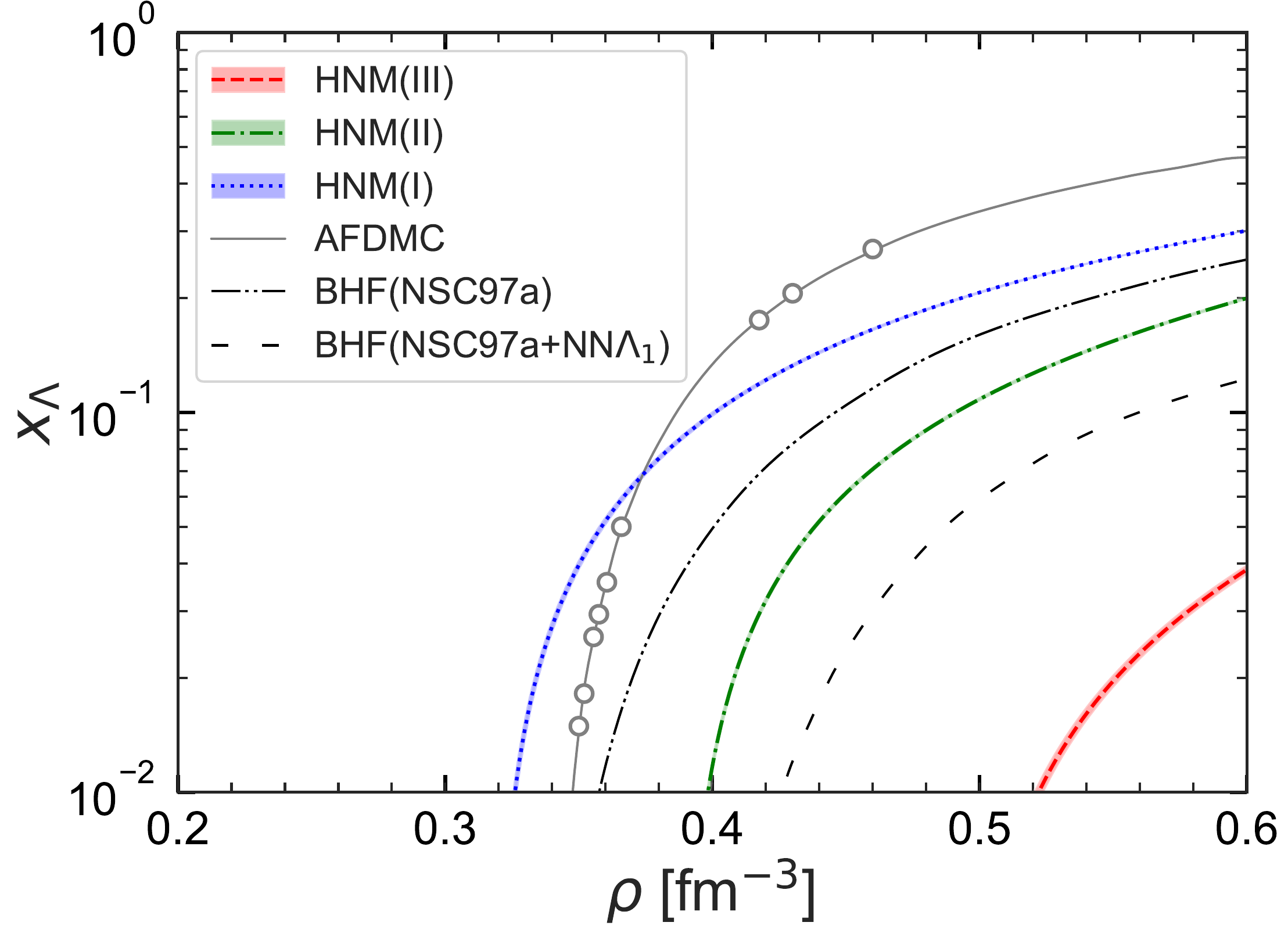}\\
  \caption{$\Lambda$-fractions for our three HNM EoSs and the one
  from AFDMC~\citep{Lonardoni:2014bwa}. The gray circles denote a different number of neutrons $(N_n = 66, 54, 38)$ and hyperons $(N_\Lambda = 1, 2, 14)$ in the simulation box giving momentum closed shells in AFDMC. The black dash-dot-dotted curve is calculated using non-relativistic BHF theory with only the $N\Lambda$ interaction, specifically Nijmegen Soft-Core 97 (NSC97). The black loosely dashed curve is calculated by including an additional $NN\Lambda$ force derived within the framework of $\chi$EFT~\citep{Logoteta:2019utx}.}
  \label{fig4}
\end{figure}

In Fig.~\ref{fig4}, $\Lambda$-fractions for our three HNM EoSs are shown.
At the given density, the Lambda fraction from the HNM~(III) is the smallest one, further indicating that the EoS for HNM~(III) is the stiffest.
In Ref.~\citep{Lonardoni:2014bwa}, they perform calculations with $N_\Lambda =1,2,14$ hyperons and use a phenomenological hyperon-nucleon potential based on the work of~\citep{Bodmer:1984gc}.
The EoS of HNM is then derived with an extrapolation function $f(\rho, x_\Lambda)$, which is quadratic in density and cubic in the $\Lambda$-fraction $x_\Lambda$.
Clearly, our calculations improve upon this by covering the full range of densities and $\Lambda$-fractions relevant to the problem at hand.
In Ref.~\citep{Lonardoni:2014bwa}, the $\Lambda$-fraction increases at higher densities under their parametrization~(I) of the $NN\Lambda$ force, predicting a maximum neutron star mass of $1.36(5)M_\odot$, as shown in Fig.~\ref{fig4}.
Under their parametrization~(II) of the $NN\Lambda$ force, the $\Lambda$-fraction drops to zero at higher densities, allowing for neutron star masses above $2M_\odot$.
For comparison, we also present another result examining the effects of the $NN\Lambda$ three-body force on neutron star properties using the BHF approach~\citep{Logoteta:2019utx}. It is notable that the strongly repulsive three-body hyperonic interactions lead to an increase in the threshold density for the $\Lambda$ hyperons and a reduction in the $\Lambda$ hyperon fraction.
It is also important to consider $\beta$-stable nuclear matter for a more realistic depiction of neutron stars, as the proton fraction can reach approximately 10\%-30\% at their cores, depending on the symmetry energy and density~\citep{Tong:2022yml, Bombaci:2018ksa}. However, incorporating protons introduces additional constraints from chemical equilibrium among protons, neutrons, electrons, and muons, as well as charge neutrality, which significantly increases the computational cost of our NLEFT simulations. For this reason, we have focused on neutrons and $\Lambda$ hyperons in the present work, and we will extend our approach to include $\beta$-stable matter in the next step.

\begin{figure}[htbp]
  \centering
  \includegraphics[width=0.47\textwidth]{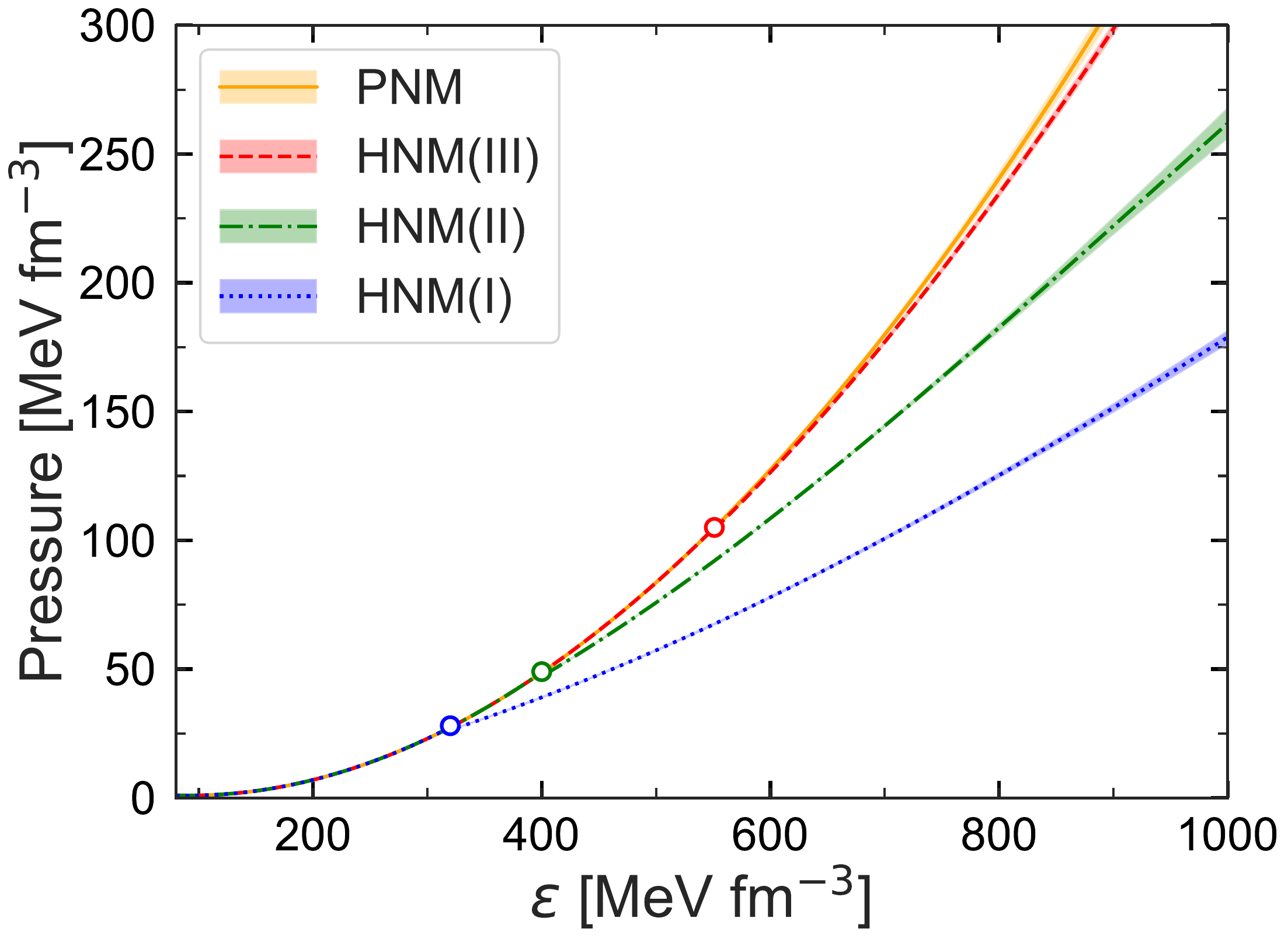}\\
  \caption{Pressure as a function of energy density. Shown are the results for PNM as well as the three HNM EoS considered here.}
  \label{fig5}
\end{figure}

In Fig.~\ref{fig5}, the neutron star EoSs for PNM and for HNM are displayed.
With increasing energy density, the pressure increases.
The threshold energy density is $\varepsilon_\Lambda^{\rm th} = 318(1)(1)$~MeV~fm$^{-3}$, $400(1)(1)$~MeV~fm$^{-3}$, and $551(1)(1)$~MeV~fm$^{-3}$ for HNM(I), HNM(II), and HNM(III), respectively.
The introduction of $\Lambda$ hyperons results in a significant softening of the EoS compared to PNM, indicating a substantial change in stiffness at higher densities.
As expected, the inclusion of $\Lambda$ hyperons softens the EoS, with HNM(III) displaying the stiffest EoS among the hyperonic cases, highlighting differences in hyperon interactions across the calculations.
This emphasizes the critical role of hyperons in influencing the stiffness and stability of neutron star matter at supra-saturation nuclear densities.

\begin{figure}[htbp]
  \centering
  \includegraphics[width=0.47\textwidth]{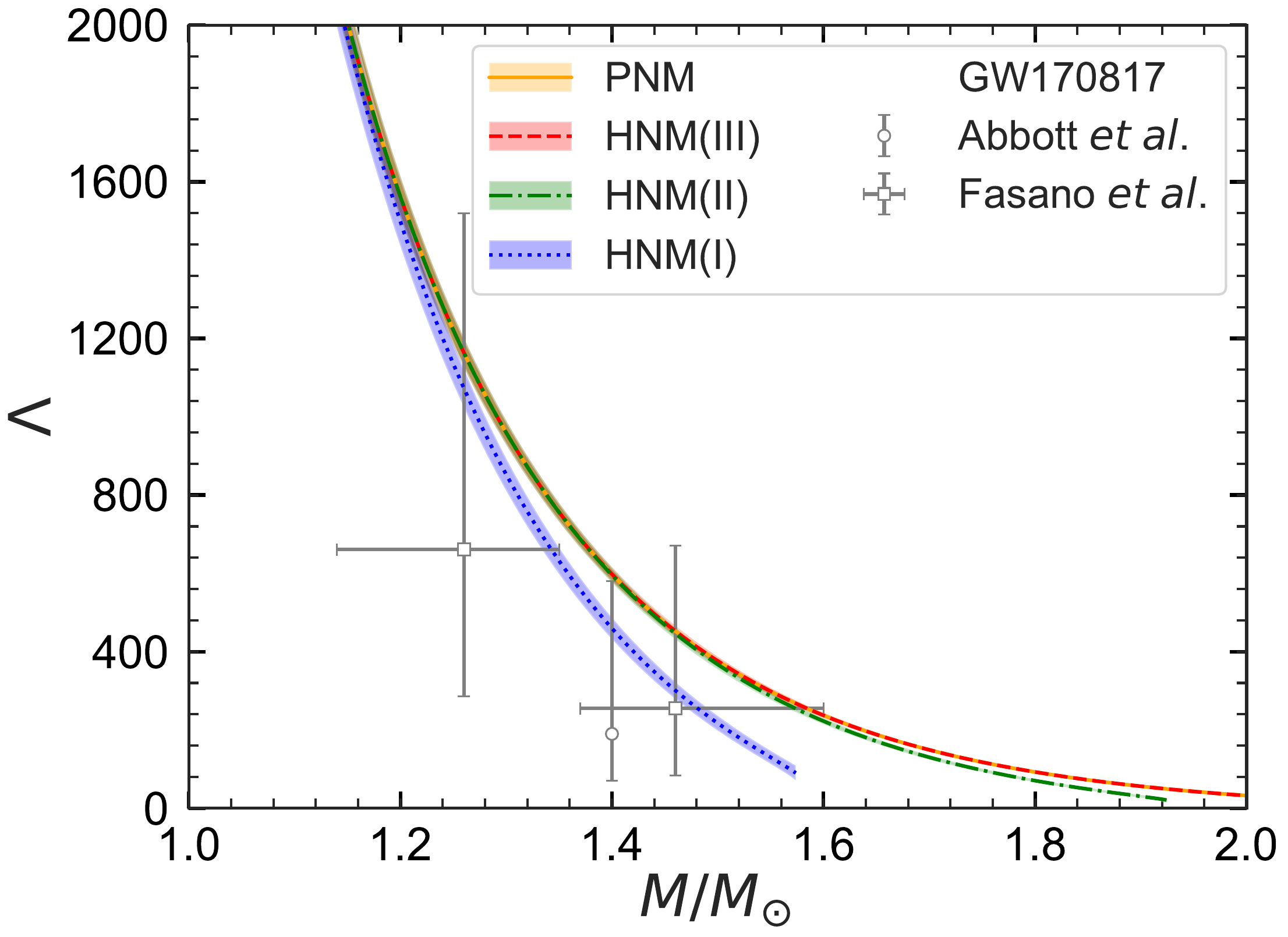}\\
  \caption{Neutron star tidal deformability, $\Lambda$, as a function of neutron star mass. $\Lambda(M)$ is compared to the masses and tidal deformabilities inferred in Ref.~\citep{Fasano:2019zwm} for the two neutron stars in the merger event GW170817 at the 90\% level (open squares) as well as $\Lambda(1.4M_\odot)$ at the 90\% level extracted from GW170817~\citep{LIGOScientific:2018cki}~(open circle).}
  \label{fig6}
\end{figure}

In the multimessenger era, another important constraint of the canonical neutron star mass (1.4$M_\odot$) is the tidal deformability $\Lambda_{1.4M_\odot}$.
In Fig.~\ref{fig6}, the tidal deformability $\Lambda_{1.4M_\odot}$ for PNM, HNM(I), HNM(II), and HNM(III) from the NLEFT are 597(5)(18), 451(5)(31), 587(5)(19), and 597(5)(18), respectively. The HNM(III) gives the largest value.
The initial estimation for the tidal deformability $\Lambda_{1.4M_\odot}$ has an upper bound $\Lambda_{1.4M_\odot}<800$~\citep{LIGOScientific:2017vwq} from the observation of Binary Neutron Star merger event GW170817.
Then a revised analysis from the LIGO and Virgo collaborations gave $\Lambda_{1.4M_\odot}=190_{-120}^{+390}$~\citep{LIGOScientific:2018cki}.
It is important to underscore that our results are located in these regions and agree well with the one inferred in Ref.~\citep{Fasano:2019zwm} for the two neutron stars in the merger event GW170817 at the 90\% level.

\begin{table*}[htbp]
    \caption{Numerical coefficients for the fit formula of the $I$-Love, $I$-$Q$, and $Q$-Love relations.}
    \label{tab:I-love-Coeff}
    \begin{tabular}{|c|c|c|c|c|c|c|}
    \hline
               $y_i$  & $x_i$ & $a_i$ & $b_i$ & $c_i$ & $d_i$ & $e_i$ \\
    \hline
    $\bar{I}$ & $\Lambda$ & $1.49081\times 10^{0}$ & $5.93228\times 10^{-2}$ & $2.25755\times 10^{-2}$ & $-7.05724\times 10^{-4}$ & $8.22849\times 10^{-6}$  \\
    $\bar{Q}$ & $\Lambda$ & $1.95541\times 10^{-1}$ & $9.42324\times 10^{-2}$ & $4.84774\times 10^{-2}$ & $-4.45415\times 10^{-3}$ & $1.35698\times 10^{-4}$ \\
    $\bar{I}$ & $\bar{Q}$ & $1.40552\times 10^{0}$ &  $5.15966\times 10^{-1}$ & $4.82729\times 10^{-2}$ & $1.69043\times 10^{-2}$ & $1.16931\times 10^{-4}$  \\
    \hline
    \end{tabular}
\end{table*}

\begin{figure}[htbp]
  \centering
  \includegraphics[width=0.45\textwidth]{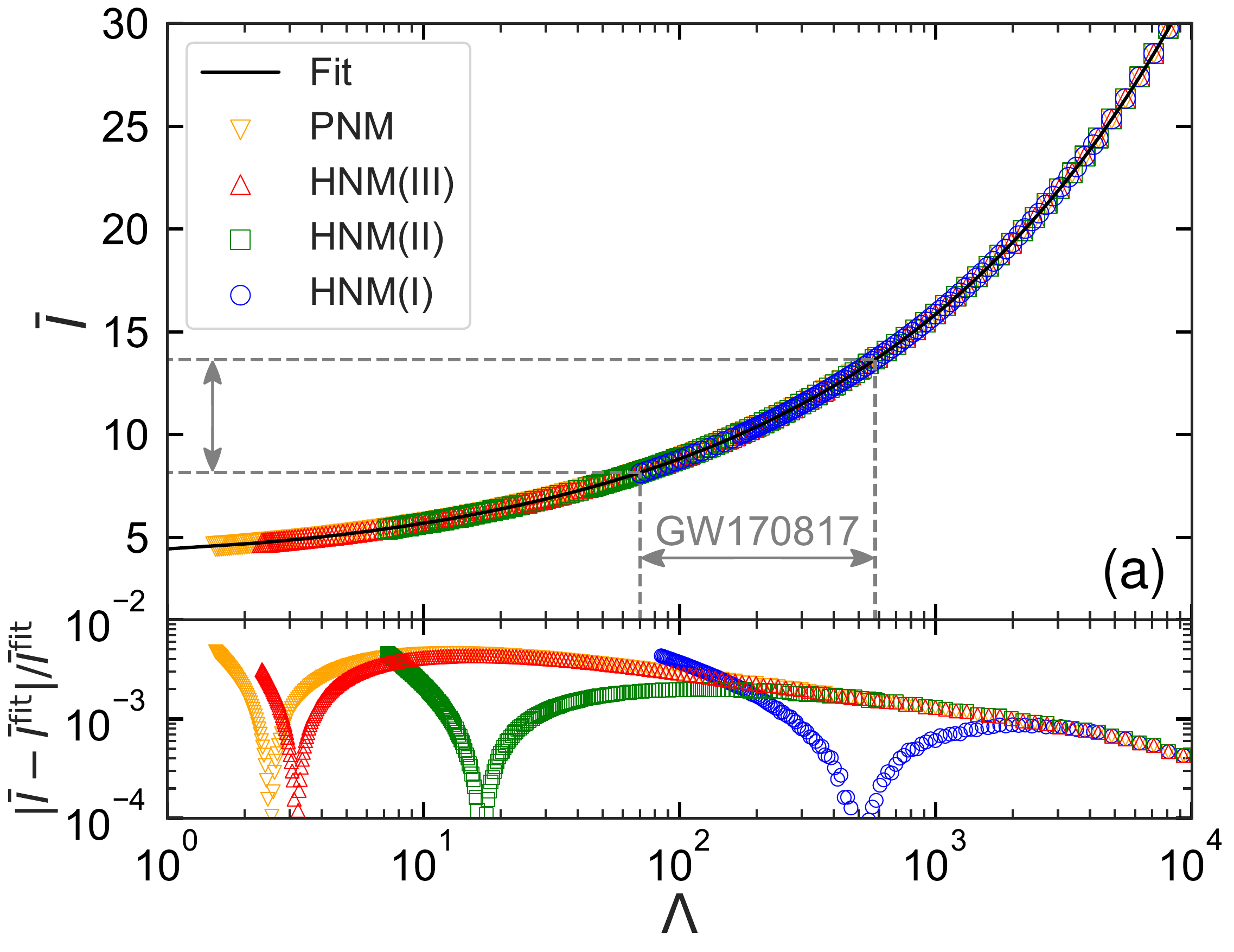}\\
  \includegraphics[width=0.45\textwidth]{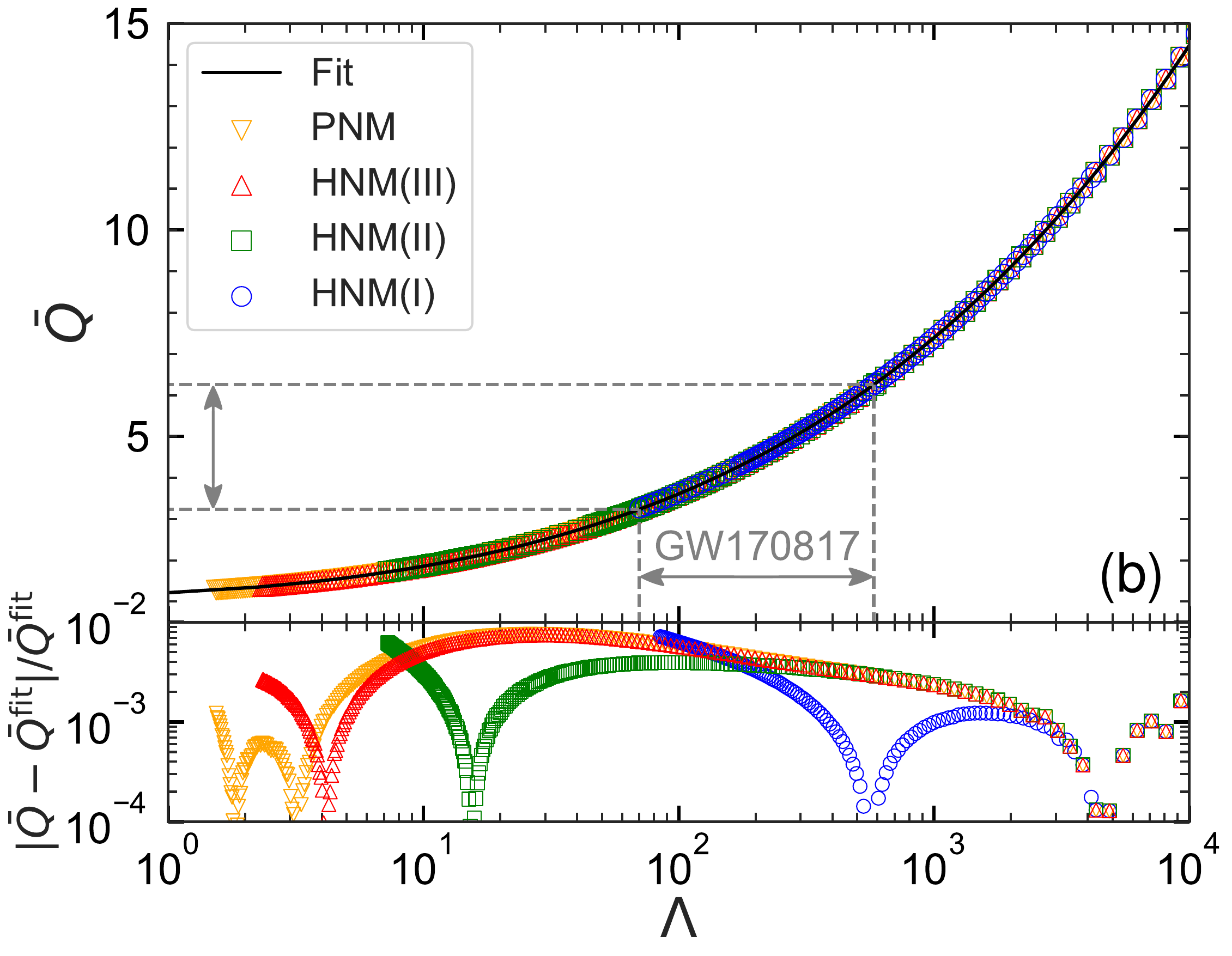}\\
  \includegraphics[width=0.45\textwidth]{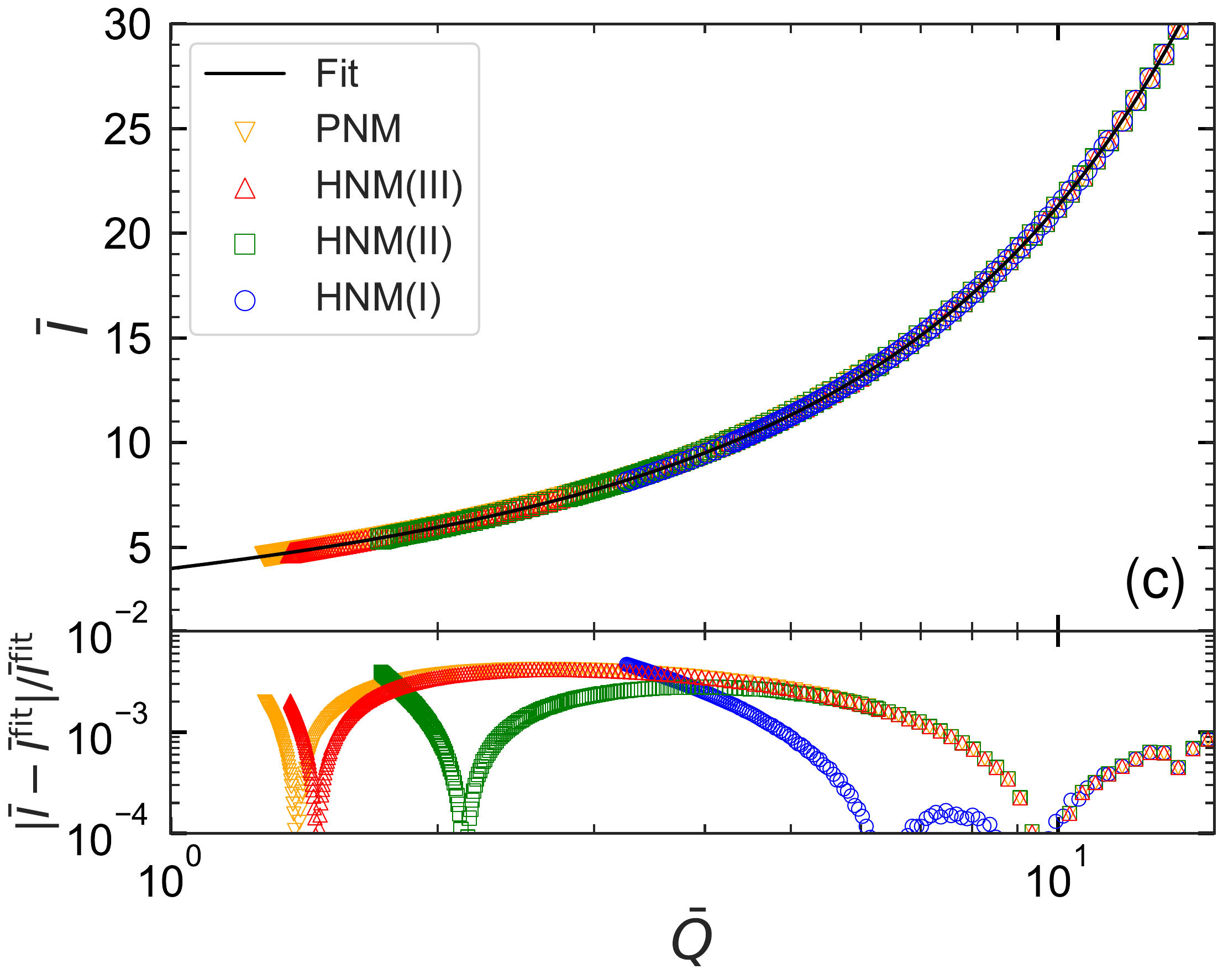}\\
  \caption{ Universal relations for PNM and HNM together within the slow-rotation approximation. The black solid line is the fitted curve, the bottom panel is the absolute fractional difference between the fit curves and the numerical results. (a) $I$-Love relation, (b) $Q$-Love relation, (c) $I$-$Q$ relation. }
  \label{fig7}
\end{figure}

The integral quantities of a neutron star, such as the mass, radius, moment of inertia, and quadrupole moment, depend sensitively on the neutron star's internal structure and thus on the EoS~\citep{Greif:2020pju}. However, the universal $I$-Love-$Q$ relations, which connect the moment of inertia $I$, tidal deformability $\Lambda$, and the quadrupole moment $Q$ in a slow rotation approximation, have been established for both hadronic EoSs and hyperonic EoSs from phenomenological approaches in recent years~\citep{Yagi:2013bca,Yagi:2013awa,Yagi:2016bkt,Sedrakian:2022ata}. The $I$-Love relations for neutron star matter with hyperons from our {\em ab initio} calculations are shown in the Fig.~\ref{fig7}(a). The dimensionless moment of inertia $\bar{I}$ is defined as $\bar{I}\equiv I/M^3$. As suggested in Refs.~\citep{Yagi:2013bca,Yagi:2013awa,Yagi:2016bkt,Sedrakian:2022ata}, the universal relations of $\bar{I}$ and $\Lambda$ can be explored by using the ansatz, $\ln y_i = a_i+b_i\ln x_i+c_i(\ln x_i)^2+d_i(\ln x_i)^3+e_i(\ln x_i)^4$, where the coefficients are listed in Table~\ref{tab:I-love-Coeff}. These coefficients closely resemble those in Ref.~\citep{Yagi:2016bkt,Li:2023owg}, where a large number of EoSs are considered. The bottom panels show the absolute fractional difference between all the data and the fit, which remains below 1\% across the entire range. Consequently, these relations are highly insensitive to whether the input EoSs include hyperons and demonstrate a high level of accuracy. While the underlying cause of this universal behavior remains incompletely understood, its practical utility is promising. By aiding in the constraint of quantities challenging to observe directly and by eliminating uncertainties related to the EoS during data analysis, it serves as a valuable tool. This universal relation enables the extraction of the moment of inertia of a neutron star with a mass of $1.4M_\odot$, denoted as $\bar{I}_{1.4M_\odot}$, from the tidal deformability $\Lambda_{1.4M_\odot}$ observed in GW170817. The revised analysis from the LIGO and Virgo Collaborations, $\Lambda_{1.4M_\odot}=190_{-120}^{+390}$~\citep{LIGOScientific:2018cki}, leads to $\bar{I}_{1.4M_\odot}=10.25_{-2.10}^{+3.40}$ as shown in Fig.~\ref{fig7}(a). These values are consistent with other results, such as $\bar{I}_{1.4M_\odot} = 11.10_{-2.28}^{+3.64}$ obtained using a large set of candidate neutron star EoSs based on relativistic mean-field and Skyrme-Hartree-Fock theory~\citep{Landry:2018jyg} and $\bar{I}_{1.4M_\odot} = 10.30^{+3.39}_{-2.10}$ from the relativistic BHF theory in the full Dirac space~\citep{Wang:2022cpi}. The $Q$-Love and $I$-$Q$ relations are also shown in Fig.~\ref{fig7}(b) and Fig.~\ref{fig7}(c).

\begin{figure*}[htbp]
  \centering
  \includegraphics[width=0.75\textwidth]{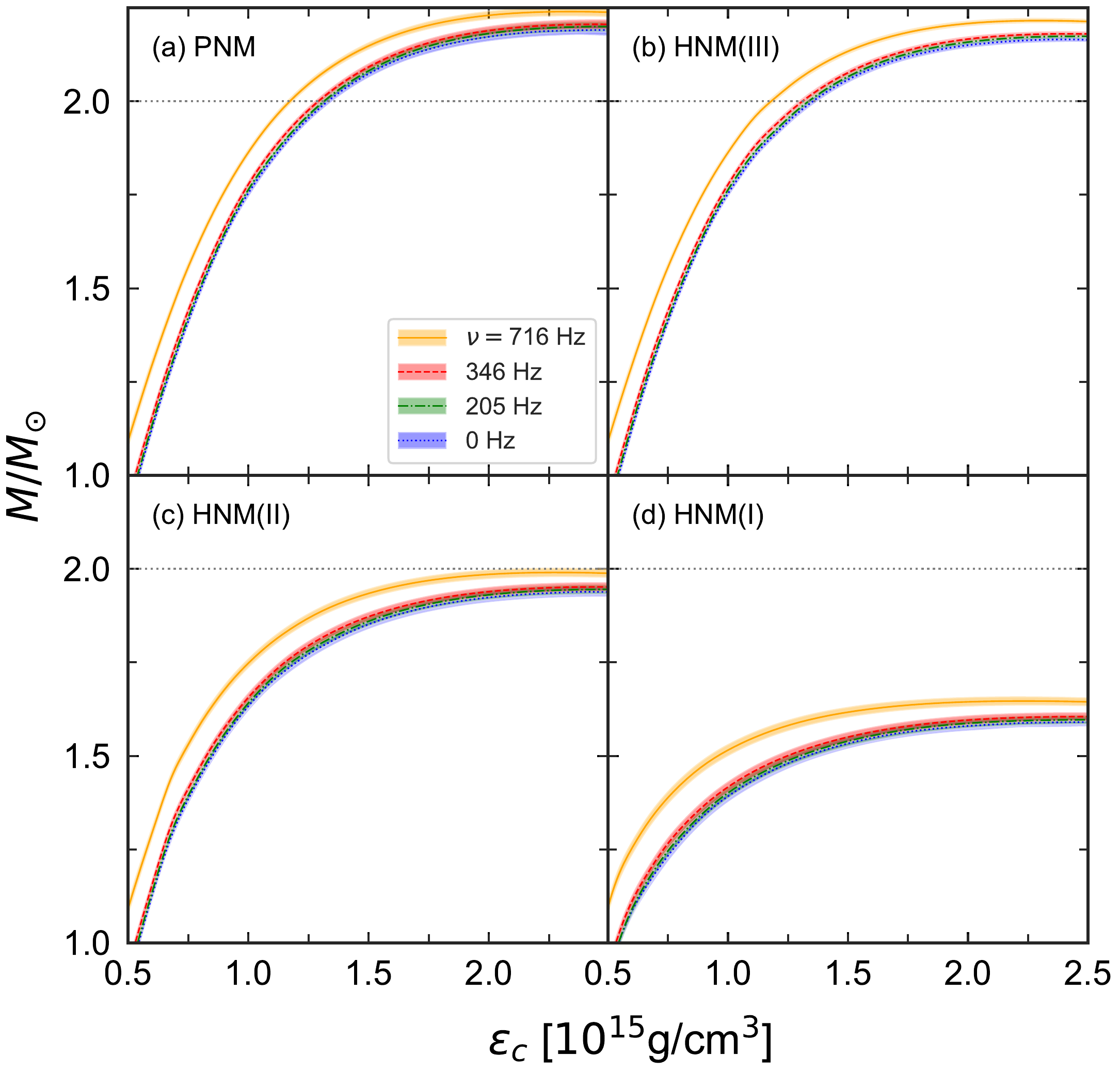}\\
  \caption{Gravitational mass $M$ as a function of central energy density $\varepsilon_c$. Four cases are shown for constant spin frequencies $\nu = 0, 205, 346, 716$ Hz. These calculations are based on the EoSs of PNM and HNM from the NLEFT calculations.}
  \label{fig8}
\end{figure*}

In addition to studying the properties of neutron stars in static and slow-rotation approximation, exploring their properties under rapid rotation is also a fascinating and significant area of research.
We evaluate the effects of uniform rotation on two millisecond pulsars observed by the NICER collaboration: PSR J0030+0451 and PSR J0740+6620, with rotational frequencies of 205~Hz~\citep{Vinciguerra2024APJ}, 346~Hz~\citep{Salmi2024arXiv} and the most rapid known pulsar PSR J1748-2446ad observed to date with 716~Hz~\citep{Hessels2006Science}.
The effect of rotation on stellar structures plays a crucial role in determining the neutron star mass at a given central energy density. Fig.~\ref{fig8} illustrates how the gravitational mass varies with central energy density for both static and rotating configurations, using the EoSs of PNM and HNM from NLEFT calculations.
At a given central energy density, the mass of a neutron star increases with increasing rotational frequency for both PNM and HNM. This underscores the significant impact of centrifugal forces on the structure of neutron stars. Also, for a given non-zero frequency and central energy density, the neutron star mass obtained from HNM(III) remains larger than those from HNM(II) and HNM(I), consistent with the conclusions drawn for static case. Specifically, for HNM(II), it can support a neutron star with $2{M_\odot}$ when the rotational frequency reaches 716 Hz.

\begin{figure*}[htbp]
  \centering
  \includegraphics[width=0.75\textwidth]{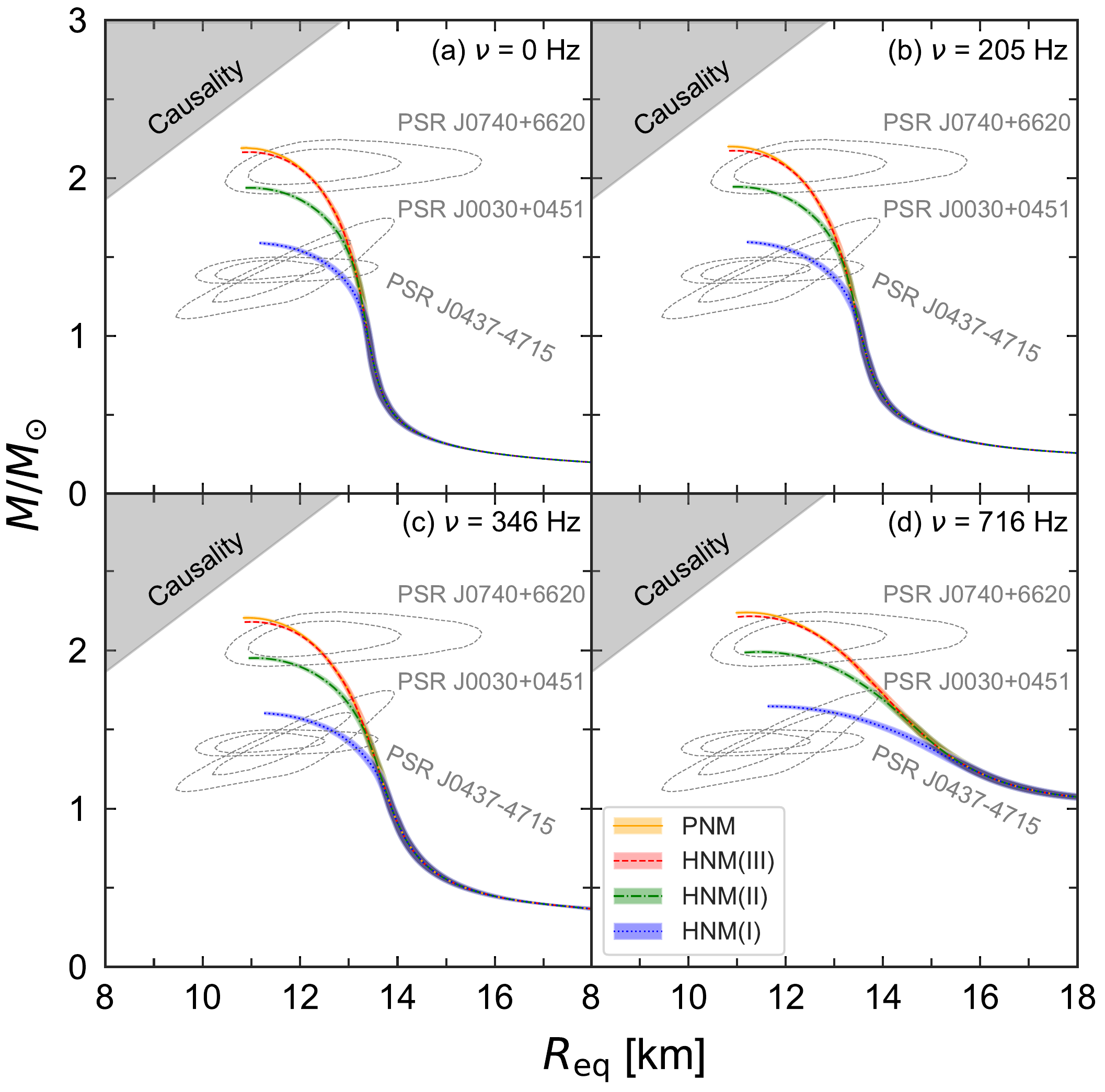}\\
  \caption{The gravitational mass $M$ as a function of the equatorial radii $R_{\rm eq}$. Four cases are shown for constant spin frequencies $\nu = 0, 205, 346, 716$ Hz. Results are obtained with EoSs of PNM and HNM from the LEFT calculations. The 68\% and 95\% credible regions of joint estimations on mass and radius for PSR J0437-4715~\citep{Choudhury2024APJ}, PSR J0740+6620~\citep{Salmi2024arXiv}, and PSR J0030+0451~\citep{Vinciguerra2024APJ} from NICER are also shown.
  The gray shaded area in the upper left corner denotes the constraints from the causality.}
  \label{fig9}
\end{figure*}

In Fig.~\ref{fig9}, the gravitational masses of both static and rotating neutron stars from PNM and HNM are plotted as functions of their equatorial radii.
The maximum masses for PNM, HNM(I), HNM(II), and HNM(III) are as follows: at $\nu=0$ Hz, 2.19(1)(1), 1.59(1)(1), 1.94(1)(1), and 2.17(1)(1)~$M_\odot$; at $\nu = 205$ Hz, 2.20(1)(1), 1.60(1)(1), 1.95(1)(1), and 2.18(1)(1)~$M_\odot$; at $\nu = 346$ Hz, 2.21(1)(1), 1.61(1)(1), 1.96(1)(1), and 2.19(1)(1)~$M_\odot$; and at $\nu = 716$ Hz, 2.24(1)(1), 1.64(1)(1), 1.99(1)(1), and 2.22(1)(1)~$M_\odot$.
It can be observed that uniform rotation at $\nu = 205$ and 346~Hz has negligible effects on the maximum mass and only a slight impact on intermediate masses. Even at the largest frequency of $\nu = 716$ Hz, the impact on the maximum mass is minimal, resulting in an increase of approximately 0.05 $M_\odot$.
In addition, the radii of a 1.4$M_\odot$ neutron star for PNM, HNM(I), HNM(II), and HNM(III) are as follows:  at $\nu=0$ Hz, $R_{1.4M\odot}=13.10(1)(7)$, $12.71(4)(13)$, $13.09(1)(8)$, and $13.10(1)(7)$~km; at $\nu = 205$ Hz, $13.26(1)(7)$, $12.89(4)(14)$, $13.25(1)(8)$, and $13.26(1)(7)$~km; at $\nu = 346$ Hz, $13.45(1)(7)$, $13.10(4)(14)$, $13.44(1)(8)$, and $13.45(1)(7)$~km; and at $\nu = 716$ Hz, $15.09(2)(14)$, $14.86(5)(15)$, $15.08(2)(14)$, and $15.09(2)(14)$~km.
Therefore, it is noteworthy that neutron stars spinning at 716~Hz exhibit a significant impact on their radii, particularly for low-mass stars, resulting in an increase of approximately 2~km. This increased rotational frequency leads to a noticeable expansion in the equatorial radii, altering the overall structure of the star. Such effects are more pronounced in lower-mass neutron stars, where the centrifugal forces induced by rapid rotation have a greater influence on the star's geometry and stability.
To further compare our results of mass-radius relations to astrophysical constraints, we also show the 68\% and 95\% credible regions of joint estimations on mass and radius for PSR J0437-4715~\citep{Choudhury2024APJ}, PSR J0740+6620~\citep{Salmi2024arXiv}, and PSR J0030+0451~\citep{Vinciguerra2024APJ} from NICER collaboration in Fig.~\ref{fig9}.
At $\nu = 0$, 205, and 346 Hz, the neutron star radii predicted by PNM, HNM(III), and HNM(II) align well with the observations of all three NICER sources, but the radii predicted by HNM (I) are consistent only with PSR J0437-4715 and PSR J0030+0451.
For low-mass neutron stars with radii below approximately 18 km in Fig.~\ref{fig9}, the spin frequency of 716 Hz remains below the Kepler frequency, ensuring stability. As the mass decreases further, the spin frequency approaches the Kepler limit, resulting in instability.

\section{Summary}\label{summary}

In summary, by utilizing a recently developed auxiliary field quantum Monte Carlo algorithm, free from sign oscillations, we derive the equation of state (EoSs) through \textit{ab initio} calculations, incorporating a significant number of hyperons. Based on these EoSs from the Nuclear Lattice Effective Field Theory, we investigate the structural properties of both non-rotating and rotating neutron stars. The analysis provides key physical quantities, including chemical potentials, particle fractions, pressure, energy density, neutron star mass, radius, tidal deformability, and the universal $I$-Love-$Q$ relation. Notably, our study explores both non-rotating and rotating configurations for neutron stars. For a given central energy density, the inclusion of rotation enables a neutron star to achieve a gravitational mass higher than the non-rotating counterpart. The most rapid spin frequency 716~Hz has a significant impact on the neutron star radii, particularly for low-mass stars, reflecting significant impact of centrifugal force that pushes the limits of mass and radius beyond those of static configurations. Similar patterns in the mass-radius relations are observed across the four different EoSs, e.g., PNM, HNM(I), HNM(II), and HNM(III), indicating that the impact of rotational dynamics on the mass-radius relation is consistent, whether the EoS includes hyperons or not. By comparing the calculated astrophysical quantities of both static and rapidly rotating neutron stars with recent astronomical observations of massive neutron stars, gravitational waves, and simultaneous mass-radius measurements, our \textit{ab initio} predictions are consistent with these observational constraints.

\section*{Acknowledgments}
\begin{acknowledgments}
We are grateful for discussions with members and partners of the Nuclear Lattice Effective Field Theory Collaboration. We are deeply thankful to Wolfram Weise for some thoughtful comments. We acknowledge funding by the European Research Council (ERC) under the European Union's Horizon 2020 research and innovation programme (AdG EXOTIC, grant agreement No. 101018170), and by the MKW
NRW under the funding code NW21-024-A.
The work of Serdar Elhatisari was further supported by  the Scientific and Technological Research Council of Turkey (TUBITAK project no. 120F341).
The work of Ulf-G.~Mei{\ss}ner was further supported by CAS through the President's International Fellowship Initiative (PIFI)
(Grant No. 2025PD0022).
\end{acknowledgments}


\bibliography{reference}{}

\begin{thebibliography}{}
\expandafter\ifx\csname natexlab\endcsname\relax\def\natexlab#1{#1}\fi
\providecommand{\url}[1]{\href{#1}{#1}}
\providecommand{\dodoi}[1]{doi:~\href{http://doi.org/#1}{\nolinkurl{#1}}}
\providecommand{\doeprint}[1]{\href{http://ascl.net/#1}{\nolinkurl{http://ascl.net/#1}}}
\providecommand{\doarXiv}[1]{\href{https://arxiv.org/abs/#1}{\nolinkurl{https://arxiv.org/abs/#1}}}

\bibitem[{Abbott {et~al.}(2017)}]{LIGOScientific:2017vwq}
Abbott, B.~P., {et~al.} 2017, Phys. Rev. Lett., 119, 161101,
  \dodoi{10.1103/PhysRevLett.119.161101}

\bibitem[{Abbott {et~al.}(2018)}]{LIGOScientific:2018cki}
---. 2018, Phys. Rev. Lett., 121, 161101,
  \dodoi{10.1103/PhysRevLett.121.161101}

\bibitem[{Alexander {et~al.}(1968)Alexander, Karshon, Shapira, Yekutieli,
  Engelmann, Filthuth, \& Lughofer}]{Alexander:1968acu}
Alexander, G., Karshon, U., Shapira, A., {et~al.} 1968, Phys. Rev., 173, 1452,
  \dodoi{10.1103/PhysRev.173.1452}

\bibitem[{Antoniadis {et~al.}(2013)}]{Antoniadis2013}
Antoniadis, J., {et~al.} 2013, Sci, 340, 1233232,
  \dodoi{10.1126/science.1233232}

\bibitem[{Arzoumanian {et~al.}(2018)}]{NANOGrav:2017wvv}
Arzoumanian, Z., {et~al.} 2018, Astrophys. J. Suppl., 235, 37,
  \dodoi{10.3847/1538-4365/aab5b0}

\bibitem[{Astashenok {et~al.}(2014)Astashenok, Capozziello, \&
  Odintsov}]{Astashenok:2014pua}
Astashenok, A.~V., Capozziello, S., \& Odintsov, S.~D. 2014, Phys. Rev. D, 89,
  103509, \dodoi{10.1103/PhysRevD.89.103509}

\bibitem[{Baym {et~al.}(1971{\natexlab{a}})Baym, Bethe, \&
  Pethick}]{Baym1971-BBP}
Baym, G., Bethe, H.~A., \& Pethick, C.~J. 1971{\natexlab{a}}, NuPhA, 175, 225 ,
  \dodoi{https://doi.org/10.1016/0375-9474(71)90281-8}

\bibitem[{Baym {et~al.}(1971{\natexlab{b}})Baym, Pethick, \&
  Sutherland}]{Baym1971-BPS}
Baym, G., Pethick, C., \& Sutherland, P. 1971{\natexlab{b}}, ApJ, 170, 299,
  \dodoi{10.1086/151216}

\bibitem[{Bodmer {et~al.}(1984)Bodmer, Usmani, \& Carlson}]{Bodmer:1984gc}
Bodmer, A.~R., Usmani, Q.~N., \& Carlson, J. 1984, Phys. Rev. C, 29, 684,
  \dodoi{10.1103/PhysRevC.29.684}

\bibitem[{Bombaci \& Logoteta(2018)}]{Bombaci:2018ksa}
Bombaci, I., \& Logoteta, D. 2018, Astron. Astrophys., 609, A128,
  \dodoi{10.1051/0004-6361/201731604}

\bibitem[{Borasoy {et~al.}(2006)Borasoy, Krebs, Lee, \&
  Mei{\ss}ner}]{Borasoy:2005yc}
Borasoy, B., Krebs, H., Lee, D., \& Mei{\ss}ner, U.-G. 2006, Nucl. Phys. A,
  768, 179, \dodoi{10.1016/j.nuclphysa.2006.01.009}

\bibitem[{Bour(2009)}]{ShahinMSc}
Bour, S. 2009, {Hyperon-Nucleon Interactions on the Lattice} (University of
  Bonn)

\bibitem[{Burgio {et~al.}(2021{\natexlab{a}})Burgio, Schulze, Vida{\~{n}}a, \&
  Wei}]{BURGIO2021PPNP}
Burgio, G., Schulze, H.-J., Vida{\~{n}}a, I., \& Wei, J.-B. 2021{\natexlab{a}},
  PrPNP, 120, 103879, \dodoi{https://doi.org/10.1016/j.ppnp.2021.103879}

\bibitem[{Burgio {et~al.}(2021{\natexlab{b}})Burgio, Schulze, Vidana, \&
  Wei}]{Burgio:2021vgk}
Burgio, G.~F., Schulze, H.~J., Vidana, I., \& Wei, J.~B. 2021{\natexlab{b}},
  Prog. Part. Nucl. Phys., 120, 103879, \dodoi{10.1016/j.ppnp.2021.103879}

\bibitem[{Chatterjee \& Vida\~na(2016)}]{Chatterjee:2015pua}
Chatterjee, D., \& Vida\~na, I. 2016, Eur. Phys. J. A, 52, 29,
  \dodoi{10.1140/epja/i2016-16029-x}

\bibitem[{Choudhury {et~al.}(2024)Choudhury, Salmi, Vinciguerra, Riley, Kini,
  Watts, Dorsman, Bogdanov, Guillot, Ray, Reardon, Remillard, Bilous,
  Huppenkothen, Lattimer, Rutherford, Arzoumanian, Gendreau, Morsink, \&
  Ho}]{Choudhury2024APJ}
Choudhury, D., Salmi, T., Vinciguerra, S., {et~al.} 2024, The Astrophysical
  Journal Letters, 971, L20, \dodoi{10.3847/2041-8213/ad5a6f}

\bibitem[{Cromartie {et~al.}(2020)}]{Cromartie2020NA}
Cromartie, H.~T., {et~al.} 2020, NatAs, 4, 72,
  \dodoi{10.1038/s41550-019-0880-2}

\bibitem[{Dapo {et~al.}(2010)Dapo, Schaefer, \& Wambach}]{Djapo:2008au}
Dapo, H., Schaefer, B.-J., \& Wambach, J. 2010, Phys. Rev. C, 81, 035803,
  \dodoi{10.1103/PhysRevC.81.035803}

\bibitem[{Demorest {et~al.}(2010)Demorest, Pennucci, Ransom, Roberts, \&
  Hessels}]{Demorest2010Nature}
Demorest, P.~B., Pennucci, T., Ransom, S.~M., Roberts, M. S.~E., \& Hessels, J.
  W.~T. 2010, Nature, 467, 1081, \dodoi{10.1038/nature09466}

\bibitem[{Elhatisari \& Lee(2014)}]{Elhatisari:2014lka}
Elhatisari, S., \& Lee, D. 2014, Phys. Rev. C, 90, 064001,
  \dodoi{10.1103/PhysRevC.90.064001}

\bibitem[{Elhatisari {et~al.}(2015)Elhatisari, Lee, Rupak, Epelbaum, Krebs,
  L\"ahde, Luu, \& Mei\ss{}ner}]{Elhatisari:2015iga}
Elhatisari, S., Lee, D., Rupak, G., {et~al.} 2015, Nature, 528, 111,
  \dodoi{10.1038/nature16067}

\bibitem[{Elhatisari {et~al.}(2016)}]{Elhatisari:2016owd}
Elhatisari, S., {et~al.} 2016, Phys. Rev. Lett., 117, 132501,
  \dodoi{10.1103/PhysRevLett.117.132501}

\bibitem[{Elhatisari {et~al.}(2024)}]{Elhatisari:2022zrb}
---. 2024, Nature, 630, 59, \dodoi{10.1038/s41586-024-07422-z}

\bibitem[{Epelbaum {et~al.}(2011)Epelbaum, Krebs, Lee, \&
  Mei{\ss}ner}]{Epelbaum:2011md}
Epelbaum, E., Krebs, H., Lee, D., \& Mei{\ss}ner, U.-G. 2011, Phys. Rev. Lett.,
  106, 192501, \dodoi{10.1103/PhysRevLett.106.192501}

\bibitem[{Fasano {et~al.}(2019)Fasano, Abdelsalhin, Maselli, \&
  Ferrari}]{Fasano:2019zwm}
Fasano, M., Abdelsalhin, T., Maselli, A., \& Ferrari, V. 2019, Phys. Rev.
  Lett., 123, 141101, \dodoi{10.1103/PhysRevLett.123.141101}

\bibitem[{Fattoyev \& Piekarewicz(2010)}]{Fattoyev:2010tb}
Fattoyev, F.~J., \& Piekarewicz, J. 2010, Phys. Rev. C, 82, 025810,
  \dodoi{10.1103/PhysRevC.82.025810}

\bibitem[{Fonseca {et~al.}(2016)}]{Fonseca2016APJ}
Fonseca, E., {et~al.} 2016, ApJ, 832, 167, \dodoi{10.3847/0004-637X/832/2/167}

\bibitem[{Fonseca {et~al.}(2021)Fonseca, Cromartie, Pennucci, Ray, Kirichenko,
  Ransom, Demorest, Stairs, Arzoumanian, Guillemot, Parthasarathy, Kerr,
  Cognard, Baker, Blumer, Brook, DeCesar, Dolch, Dong, Ferrara, Fiore,
  Garver-Daniels, Good, Jennings, Jones, Kaspi, Lam, Lorimer, Luo, McEwen,
  McKee, McLaughlin, McMann, Meyers, Naidu, Ng, Nice, Pol, Radovan,
  Shapiro-Albert, Tan, Tendulkar, Swiggum, Wahl, \& Zhu}]{Fonseca2021APJ}
Fonseca, E., Cromartie, H.~T., Pennucci, T.~T., {et~al.} 2021, The
  Astrophysical Journal Letters, 915, L12, \dodoi{10.3847/2041-8213/ac03b8}

\bibitem[{Frame {et~al.}(2020)Frame, L\"ahde, Lee, \&
  Mei\ss{}ner}]{Frame:2020mvv}
Frame, D., L\"ahde, T.~A., Lee, D., \& Mei\ss{}ner, U.-G. 2020, Eur. Phys. J.
  A, 56, 248, \dodoi{10.1140/epja/s10050-020-00257-y}

\bibitem[{Friedman \& Gal(2023)}]{Friedman:2022bpw}
Friedman, E., \& Gal, A. 2023, Phys. Lett. B, 837, 137669,
  \dodoi{10.1016/j.physletb.2023.137669}

\bibitem[{Gal {et~al.}(2016)Gal, Hungerford, \& Millener}]{Gal:2016boi}
Gal, A., Hungerford, E.~V., \& Millener, D.~J. 2016, Rev. Mod. Phys., 88,
  035004, \dodoi{10.1103/RevModPhys.88.035004}

\bibitem[{Gerstung {et~al.}(2020)Gerstung, Kaiser, \& Weise}]{Gerstung:2020ktv}
Gerstung, D., Kaiser, N., \& Weise, W. 2020, Eur. Phys. J. A, 56, 175,
  \dodoi{10.1140/epja/s10050-020-00180-2}

\bibitem[{Glendenning {et~al.}(1997)Glendenning, Pei, \&
  Weber}]{Glendenning1997PRL}
Glendenning, N.~K., Pei, S., \& Weber, F. 1997, Physical Review Letters, 79,
  1603, \dodoi{10.1103/PhysRevLett.79.1603}

\bibitem[{Greif {et~al.}(2020)Greif, Hebeler, Lattimer, Pethick, \&
  Schwenk}]{Greif:2020pju}
Greif, S.~K., Hebeler, K., Lattimer, J.~M., Pethick, C.~J., \& Schwenk, A.
  2020, Astrophys. J., 901, 155, \dodoi{10.3847/1538-4357/abaf55}

\bibitem[{Haidenbauer {et~al.}(2017)Haidenbauer, Mei\ss{}ner, Kaiser, \&
  Weise}]{Haidenbauer:2016vfq}
Haidenbauer, J., Mei\ss{}ner, U.-G., Kaiser, N., \& Weise, W. 2017, Eur. Phys.
  J. A, 53, 121, \dodoi{10.1140/epja/i2017-12316-4}

\bibitem[{Haidenbauer {et~al.}(2020)Haidenbauer, Mei\ss{}ner, \&
  Nogga}]{Haidenbauer:2019boi}
Haidenbauer, J., Mei\ss{}ner, U.~G., \& Nogga, A. 2020, Eur. Phys. J. A, 56,
  91, \dodoi{10.1140/epja/s10050-020-00100-4}

\bibitem[{Haidenbauer {et~al.}(2016)Haidenbauer, Mei\ss{}ner, \&
  Petschauer}]{Haidenbauer:2015zqb}
Haidenbauer, J., Mei\ss{}ner, U.-G., \& Petschauer, S. 2016, Nucl. Phys. A,
  954, 273, \dodoi{10.1016/j.nuclphysa.2016.01.006}

\bibitem[{Haidenbauer {et~al.}(2013)Haidenbauer, Petschauer, Kaiser,
  Mei{\ss}ner, Nogga, \& Weise}]{Haidenbauer:2013oca}
Haidenbauer, J., Petschauer, S., Kaiser, N., {et~al.} 2013, Nucl. Phys. A, 915,
  24, \dodoi{10.1016/j.nuclphysa.2013.06.008}

\bibitem[{Hartle(1967)}]{Hartle:1967he}
Hartle, J.~B. 1967, Astrophys. J., 150, 1005, \dodoi{10.1086/149400}

\bibitem[{Hartle \& Thorne(1968)}]{Hartle:1968si}
Hartle, J.~B., \& Thorne, K.~S. 1968, Astrophys. J., 153, 807,
  \dodoi{10.1086/149707}

\bibitem[{Hauptman {et~al.}(1977)Hauptman, Kadyk, \&
  Trilling}]{Hauptman:1977hr}
Hauptman, J.~M., Kadyk, J.~A., \& Trilling, G.~H. 1977, Nucl. Phys. B, 125, 29,
  \dodoi{10.1016/0550-3213(77)90222-X}

\bibitem[{Hessels {et~al.}(2006)Hessels, Ransom, Stairs, Freire, Kaspi, \&
  Camilo}]{Hessels2006Science}
Hessels, J. W.~T., Ransom, S.~M., Stairs, I.~H., {et~al.} 2006, Science, 311,
  1901, \dodoi{doi:10.1126/science.1123430}

\bibitem[{Hildenbrand {et~al.}(2022)Hildenbrand, Elhatisari, L\"ahde, Lee, \&
  Mei\ss{}ner}]{Hildenbrand:2022imw}
Hildenbrand, F., Elhatisari, S., L\"ahde, T.~A., Lee, D., \& Mei\ss{}ner, U.-G.
  2022, Eur. Phys. J. A, 58, 167, \dodoi{10.1140/epja/s10050-022-00821-8}

\bibitem[{Hinderer(2008)}]{Hinderer:2007mb}
Hinderer, T. 2008, Astrophys. J., 677, 1216, \dodoi{10.1086/533487}

\bibitem[{Hinderer {et~al.}(2010)Hinderer, Lackey, Lang, \&
  Read}]{Hinderer:2009ca}
Hinderer, T., Lackey, B.~D., Lang, R.~N., \& Read, J.~S. 2010, Phys. Rev. D,
  81, 123016, \dodoi{10.1103/PhysRevD.81.123016}

\bibitem[{Huth {et~al.}(2022)}]{Huth:2021bsp}
Huth, S., {et~al.} 2022, Nature, 606, 276, \dodoi{10.1038/s41586-022-04750-w}

\bibitem[{Kadyk {et~al.}(1971)Kadyk, Alexander, Chan, Gaposchkin, \&
  Trilling}]{Kadyk:1971tc}
Kadyk, J.~A., Alexander, G., Chan, J.~H., Gaposchkin, P., \& Trilling, G.~H.
  1971, Nucl. Phys. B, 27, 13, \dodoi{10.1016/0550-3213(71)90076-9}

\bibitem[{Komatsu {et~al.}(1989)Komatsu, Eriguchi, \&
  Hachisu}]{Komatsu1989MNRAS}
Komatsu, H., Eriguchi, Y., \& Hachisu, I. 1989, Monthly Notices of the Royal
  Astronomical Society, 237, 355, \dodoi{10.1093/mnras/237.2.355}

\bibitem[{K\"onig {et~al.}(2017)K\"onig, Grie\ss{}hammer, Hammer, \& van
  Kolck}]{Konig:2016utl}
K\"onig, S., Grie\ss{}hammer, H.~W., Hammer, H.~W., \& van Kolck, U. 2017,
  Phys. Rev. Lett., 118, 202501, \dodoi{10.1103/PhysRevLett.118.202501}

\bibitem[{Krastev \& Sammarruca(2006)}]{Krastev:2006ii}
Krastev, P.~G., \& Sammarruca, F. 2006, Phys. Rev. C, 74, 025808,
  \dodoi{10.1103/PhysRevC.74.025808}

\bibitem[{Kumar {et~al.}(2024)}]{MUSES:2023hyz}
Kumar, R., {et~al.} 2024, Living Rev. Rel., 27, 3,
  \dodoi{10.1007/s41114-024-00049-6}

\bibitem[{L\"ahde \& Mei\ss{}ner(2019)}]{Lahde:2019npb}
L\"ahde, T.~A., \& Mei\ss{}ner, U.-G. 2019, {Nuclear Lattice Effective Field
  Theory}: {An introduction}, Vol. 957 (Springer),
  \dodoi{10.1007/978-3-030-14189-9}

\bibitem[{Landry \& Kumar(2018)}]{Landry:2018jyg}
Landry, P., \& Kumar, B. 2018, Astrophys. J. Lett., 868, L22,
  \dodoi{10.3847/2041-8213/aaee76}

\bibitem[{Lattimer \& Prakash(2004)}]{Lattimer:2004pg}
Lattimer, J.~M., \& Prakash, M. 2004, Science, 304, 536,
  \dodoi{10.1126/science.1090720}

\bibitem[{Lee(2009)}]{Lee:2008fa}
Lee, D. 2009, Prog. Part. Nucl. Phys., 63, 117,
  \dodoi{10.1016/j.ppnp.2008.12.001}

\bibitem[{Lee \& Sch\"afer(2005)}]{Lee:2004qd}
Lee, D., \& Sch\"afer, T. 2005, Phys. Rev. C, 72, 024006,
  \dodoi{10.1103/PhysRevC.72.024006}

\bibitem[{Li {et~al.}(2016)Li, Zhang, Zhang, Gao, Qi, \& Liu}]{Li2016PRD}
Li, A., Zhang, B., Zhang, N.-B., {et~al.} 2016, Phys. Rev. D, 94, 083010,
  \dodoi{10.1103/PhysRevD.94.083010}

\bibitem[{Li {et~al.}(2023)Li, Sedrakian, \& Weber}]{Li:2023owg}
Li, J.~J., Sedrakian, A., \& Weber, F. 2023, Phys. Rev. C, 108, 025810,
  \dodoi{10.1103/PhysRevC.108.025810}

\bibitem[{Li {et~al.}(2019)Li, Elhatisari, Epelbaum, Lee, Lu, \&
  Mei\ss{}ner}]{Li:2019ldq}
Li, N., Elhatisari, S., Epelbaum, E., {et~al.} 2019, Phys. Rev. C, 99, 064001,
  \dodoi{10.1103/PhysRevC.99.064001}

\bibitem[{Li {et~al.}(2018)Li, Elhatisari, Epelbaum, Lee, Lu, \&
  Mei\ss{}ner}]{Li:2018ymw}
---. 2018, Phys. Rev. C, 98, 044002, \dodoi{10.1103/PhysRevC.98.044002}

\bibitem[{Logoteta {et~al.}(2019)Logoteta, Vidana, \&
  Bombaci}]{Logoteta:2019utx}
Logoteta, D., Vidana, I., \& Bombaci, I. 2019, Eur. Phys. J. A, 55, 207,
  \dodoi{10.1140/epja/i2019-12909-9}

\bibitem[{Lonardoni {et~al.}(2015)Lonardoni, Lovato, Gandolfi, \&
  Pederiva}]{Lonardoni:2014bwa}
Lonardoni, D., Lovato, A., Gandolfi, S., \& Pederiva, F. 2015, Phys. Rev.
  Lett., 114, 092301, \dodoi{10.1103/PhysRevLett.114.092301}

\bibitem[{Lovato {et~al.}(2022)Lovato, Bombaci, Logoteta, Piarulli, \&
  Wiringa}]{Lovato:2022apd}
Lovato, A., Bombaci, I., Logoteta, D., Piarulli, M., \& Wiringa, R.~B. 2022,
  Phys. Rev. C, 105, 055808, \dodoi{10.1103/PhysRevC.105.055808}

\bibitem[{Lu {et~al.}(2020)Lu, Li, Elhatisari, Lee, Drut, L\"ahde, Epelbaum, \&
  Mei\ss{}ner}]{Lu:2019nbg}
Lu, B.-N., Li, N., Elhatisari, S., {et~al.} 2020, Phys. Rev. Lett., 125,
  192502, \dodoi{10.1103/PhysRevLett.125.192502}

\bibitem[{Lu {et~al.}(2019)Lu, Li, Elhatisari, Lee, Epelbaum, \&
  Mei\ss{}ner}]{Lu:2018bat}
---. 2019, Phys. Lett. B, 797, 134863, \dodoi{10.1016/j.physletb.2019.134863}

\bibitem[{Marino {et~al.}(2024)Marino, Dehman, Kovlakas, Rea, Pons, \&
  Vigan\`o}]{Marino:2024gpm}
Marino, A., Dehman, C., Kovlakas, K., {et~al.} 2024, Nature Astron., 8.
\newblock \doarXiv{2404.05371}

\bibitem[{Maslov {et~al.}(2015)Maslov, Kolomeitsev, \&
  Voskresensky}]{Maslov:2015msa}
Maslov, K.~A., Kolomeitsev, E.~E., \& Voskresensky, D.~N. 2015, Phys. Lett. B,
  748, 369, \dodoi{10.1016/j.physletb.2015.07.032}

\bibitem[{Masuda {et~al.}(2016)Masuda, Hatsuda, \& Takatsuka}]{Masuda:2015kha}
Masuda, K., Hatsuda, T., \& Takatsuka, T. 2016, Eur. Phys. J. A, 52, 65,
  \dodoi{10.1140/epja/i2016-16065-6}

\bibitem[{Mei\ss{}ner {et~al.}(2024)Mei\ss{}ner, Shen, Elhatisari, \&
  Lee}]{Meissner:2023cvo}
Mei\ss{}ner, U.-G., Shen, S., Elhatisari, S., \& Lee, D. 2024, Phys. Rev.
  Lett., 132, 062501, \dodoi{10.1103/PhysRevLett.132.062501}

\bibitem[{Oertel {et~al.}(2017)Oertel, Hempel, Kl\"ahn, \& Typel}]{Oertel2017}
Oertel, M., Hempel, M., Kl\"ahn, T., \& Typel, S. 2017, RvMP, 89, 015007,
  \dodoi{10.1103/RevModPhys.89.015007}

\bibitem[{Oppenheimer \& Volkoff(1939)}]{Oppenheimer:1939ne}
Oppenheimer, J.~R., \& Volkoff, G.~M. 1939, Phys. Rev., 55, 374,
  \dodoi{10.1103/PhysRev.55.374}

\bibitem[{Paschalidis \& Stergioulas(2017)}]{Paschalidis2017LR}
Paschalidis, V., \& Stergioulas, N. 2017, Living Reviews in Relativity, 20, 7,
  \dodoi{10.1007/s41114-017-0008-x}

\bibitem[{Petschauer {et~al.}(2017)Petschauer, Haidenbauer, Kaiser,
  Mei\ss{}ner, \& Weise}]{Petschauer:2016pbn}
Petschauer, S., Haidenbauer, J., Kaiser, N., Mei\ss{}ner, U.-G., \& Weise, W.
  2017, Nucl. Phys. A, 957, 347, \dodoi{10.1016/j.nuclphysa.2016.09.010}

\bibitem[{Petschauer {et~al.}(2016)Petschauer, Kaiser, Haidenbauer,
  Mei\ss{}ner, \& Weise}]{Petschauer:2015elq}
Petschauer, S., Kaiser, N., Haidenbauer, J., Mei\ss{}ner, U.-G., \& Weise, W.
  2016, Phys. Rev. C, 93, 014001, \dodoi{10.1103/PhysRevC.93.014001}

\bibitem[{Qu {et~al.}(2025)Qu, Wang, \& Tong}]{Qu:2024duu}
Qu, X., Wang, S., \& Tong, H. 2025, Astrophys. J., 980, 3,
  \dodoi{10.3847/1538-4357/ada76b}

\bibitem[{Ren {et~al.}(2024)Ren, Elhatisari, L\"ahde, Lee, \&
  Mei\ss{}ner}]{Ren:2023ued}
Ren, Z., Elhatisari, S., L\"ahde, T.~A., Lee, D., \& Mei\ss{}ner, U.-G. 2024,
  Phys. Lett. B, 850, 138463, \dodoi{10.1016/j.physletb.2024.138463}

\bibitem[{Salmi {et~al.}(2024)Salmi, Choudhury, Kini, Riley, Vinciguerra,
  Watts, Wolff, Arzoumanian, Bogdanov, Chakrabarty, Gendreau, Guillot, Ho,
  Huppenkothen, Ludlam, Morsink, \& Ray}]{Salmi2024arXiv}
Salmi, T., Choudhury, D., Kini, Y., {et~al.} 2024, The Radius of the High Mass
  Pulsar PSR J0740+6620 With 3.6 Years of NICER Data.
\newblock \doarXiv{2406.14466}

\bibitem[{Schaffner-Bielich(2008)}]{Schaffner-Bielich:2008zws}
Schaffner-Bielich, J. 2008, Nucl. Phys. A, 804, 309,
  \dodoi{10.1016/j.nuclphysa.2008.01.005}

\bibitem[{Schulze \& Rijken(2011)}]{Schulze:2011zza}
Schulze, H.~J., \& Rijken, T. 2011, Phys. Rev. C, 84, 035801,
  \dodoi{10.1103/PhysRevC.84.035801}

\bibitem[{Sechi-Zorn {et~al.}(1968)Sechi-Zorn, Kehoe, Twitty, \&
  Burnstein}]{Sechi-Zorn:1968mao}
Sechi-Zorn, B., Kehoe, B., Twitty, J., \& Burnstein, R.~A. 1968, Phys. Rev.,
  175, 1735, \dodoi{10.1103/PhysRev.175.1735}

\bibitem[{Sedrakian {et~al.}(2023{\natexlab{a}})Sedrakian, Li, \&
  Weber}]{Sedrakian2023PPNP}
Sedrakian, A., Li, J.~J., \& Weber, F. 2023{\natexlab{a}}, Progress in Particle
  and Nuclear Physics, 131, 104041,
  \dodoi{https://doi.org/10.1016/j.ppnp.2023.104041}

\bibitem[{Sedrakian {et~al.}(2023{\natexlab{b}})Sedrakian, Li, \&
  Weber}]{Sedrakian:2022ata}
Sedrakian, A., Li, J.-J., \& Weber, F. 2023{\natexlab{b}}, Prog. Part. Nucl.
  Phys., 131, 104041, \dodoi{10.1016/j.ppnp.2023.104041}

\bibitem[{Shen {et~al.}(2023)Shen, Elhatisari, L\"ahde, Lee, Lu, \&
  Mei\ss{}ner}]{Shen:2022bak}
Shen, S., Elhatisari, S., L\"ahde, T.~A., {et~al.} 2023, Nature Commun., 14,
  2777, \dodoi{10.1038/s41467-023-38391-y}

\bibitem[{Shen {et~al.}(2021)Shen, L\"ahde, Lee, \& Mei\ss{}ner}]{Shen:2021kqr}
Shen, S., L\"ahde, T.~A., Lee, D., \& Mei\ss{}ner, U.-G. 2021, Eur. Phys. J. A,
  57, 276, \dodoi{10.1140/epja/s10050-021-00586-6}

\bibitem[{{Stergioulas} \& {Friedman}(1995)}]{Stergioulas1995APJ}
{Stergioulas}, N., \& {Friedman}, J.~L. 1995, \apj, 444, 306,
  \dodoi{10.1086/175605}

\bibitem[{Takatsuka {et~al.}(2008)Takatsuka, Nishizaki, \&
  Tamagaki}]{Takatsuka:2008zz}
Takatsuka, T., Nishizaki, S., \& Tamagaki, R. 2008, Prog. Theor. Phys. Suppl.,
  174, 80, \dodoi{10.1143/PTPS.174.80}

\bibitem[{Tolman(1939)}]{Tolman:1939jz}
Tolman, R.~C. 1939, Phys. Rev., 55, 364, \dodoi{10.1103/PhysRev.55.364}

\bibitem[{Tong {et~al.}(2025)Tong, Elhatisari, \& Mei{\ss}ner}]{TONG2025}
Tong, H., Elhatisari, S., \& Mei{\ss}ner, U.-G. 2025, Science Bulletin,
  \dodoi{https://doi.org/10.1016/j.scib.2025.01.008}

\bibitem[{Tong {et~al.}(2022)Tong, Wang, \& Wang}]{Tong:2022yml}
Tong, H., Wang, C., \& Wang, S. 2022, Astrophys. J., 930, 137,
  \dodoi{10.3847/1538-4357/ac65fc}

\bibitem[{Tong {et~al.}(2020)Tong, Zhao, \& Meng}]{Tong2020PRC}
Tong, H., Zhao, P., \& Meng, J. 2020, PhRvC, 101, 035802,
  \dodoi{10.1103/PhysRevC.101.035802}

\bibitem[{Tsang {et~al.}(2024)Tsang, Tsang, Lynch, Kumar, \&
  Horowitz}]{Tsang:2023vhh}
Tsang, C.~Y., Tsang, M.~B., Lynch, W.~G., Kumar, R., \& Horowitz, C.~J. 2024,
  Nature Astron., 8, 328, \dodoi{10.1038/s41550-023-02161-z}

\bibitem[{Vidana {et~al.}(2011)Vidana, Logoteta, Providencia, Polls, \&
  Bombaci}]{Vidana:2010ip}
Vidana, I., Logoteta, D., Providencia, C., Polls, A., \& Bombaci, I. 2011, EPL,
  94, 11002, \dodoi{10.1209/0295-5075/94/11002}

\bibitem[{Vinciguerra {et~al.}(2024)Vinciguerra, Salmi, Watts, Choudhury,
  Riley, Ray, Bogdanov, Kini, Guillot, Chakrabarty, Ho, Huppenkothen, Morsink,
  Wadiasingh, \& Wolff}]{Vinciguerra2024APJ}
Vinciguerra, S., Salmi, T., Watts, A.~L., {et~al.} 2024, The Astrophysical
  Journal, 961, 62, \dodoi{10.3847/1538-4357/acfb83}

\bibitem[{Wang {et~al.}(2022)Wang, Wang, \& Tong}]{Wang:2022cpi}
Wang, S., Wang, C., \& Tong, H. 2022, Phys. Rev. C, 106, 045804,
  \dodoi{10.1103/PhysRevC.106.045804}

\bibitem[{Weber \& Glendenning(1991)}]{Weber1991PLB}
Weber, F., \& Glendenning, N.~K. 1991, Physics Letters B, 265, 1,
  \dodoi{https://doi.org/10.1016/0370-2693(91)90002-8}

\bibitem[{Weissenborn {et~al.}(2012)Weissenborn, Chatterjee, \&
  Schaffner-Bielich}]{Weissenborn:2011ut}
Weissenborn, S., Chatterjee, D., \& Schaffner-Bielich, J. 2012, Phys. Rev. C,
  85, 065802, \dodoi{10.1103/PhysRevC.85.065802}

\bibitem[{Wigner(1937)}]{Wigner:1936dx}
Wigner, E. 1937, Phys. Rev., 51, 106, \dodoi{10.1103/PhysRev.51.106}

\bibitem[{Yagi \& Yunes(2013{\natexlab{a}})}]{Yagi:2013awa}
Yagi, K., \& Yunes, N. 2013{\natexlab{a}}, Phys. Rev. D, 88, 023009,
  \dodoi{10.1103/PhysRevD.88.023009}

\bibitem[{Yagi \& Yunes(2013{\natexlab{b}})}]{Yagi:2013bca}
---. 2013{\natexlab{b}}, Science, 341, 365, \dodoi{10.1126/science.1236462}

\bibitem[{Yagi \& Yunes(2017)}]{Yagi:2016bkt}
---. 2017, Phys. Rept., 681, 1, \dodoi{10.1016/j.physrep.2017.03.002}

\bibitem[{Yamamoto {et~al.}(2013)Yamamoto, Furumoto, Yasutake, \&
  Rijken}]{Yamamoto:2013ada}
Yamamoto, Y., Furumoto, T., Yasutake, N., \& Rijken, T.~A. 2013, Phys. Rev. C,
  88, 022801(R), \dodoi{10.1103/PhysRevC.88.022801}

\end{thebibliography}
\bibliographystyle{aasjournal}

\end{CJK*}
\end{document}